\input harvmac
\noblackbox

\def\ev#1{\langle#1\rangle}
\def\CL {{\cal L}}
\def\CM {{\cal M}}
\def\CN {{\cal N}}
\def\unit{\relax{\rm 1\kern-.26em I}}
\def\bar{\overline}
\def\tilde{\widetilde}

\lref\GiudiceBP{
  G.~F.~Giudice and R.~Rattazzi,
  ``Theories with gauge-mediated supersymmetry breaking,''
  Phys.\ Rept.\  {\bf 322}, 419 (1999)
  [arXiv:hep-ph/9801271].
}
\lref\MurayamaFE{
  H.~Murayama and Y.~Nomura,
  ``Simple Scheme for Gauge Mediation,''
  arXiv:hep-ph/0701231.
}
\lref\MurayamaNG{
  H.~Murayama,
  ``Studying noncalculable models of dynamical supersymmetry breaking,''
  Phys.\ Lett.\ B {\bf 355}, 187 (1995)
  [arXiv:hep-th/9505082].
}
\lref\GatesNR{
S.~J.~Gates, M.~T.~Grisaru, M.~Rocek and W.~Siegel,
``Superspace, or one thousand and one lessons in supersymmetry,''
Front.\ Phys.\ {\bf 58}, 1 (1983)
[arXiv:hep-th/0108200].
}
\lref\WittenKV{
  E.~Witten,
  ``Mass Hierarchies In Supersymmetric Theories,''
  Phys.\ Lett.\ B {\bf 105}, 267 (1981).
}
\lref\WittenDF{
  E.~Witten,
  ``Constraints On Supersymmetry Breaking,''
  Nucl.\ Phys.\ B {\bf 202}, 253 (1982).
}
\lref\PeskinQI{
  M.~E.~Peskin,
  ``Duality in supersymmetric Yang-Mills theory,''
  arXiv:hep-th/9702094.
}
\lref\StrasslerQG{
  M.~J.~Strassler,
  ``An unorthodox introduction to supersymmetric gauge theory,''
  arXiv:hep-th/0309149.
}
\lref\ChackoSI{
  Z.~Chacko, M.~A.~Luty and E.~Ponton,
  ``Calculable dynamical supersymmetry breaking on deformed moduli spaces,''
  JHEP {\bf 9812}, 016 (1998)
  [arXiv:hep-th/9810253].
}
\lref\IntriligatorFK{
  K.~A.~Intriligator and S.~D.~Thomas,
  ``Dual descriptions of supersymmetry breaking,''
  arXiv:hep-th/9608046.
}
\lref\AbelCR{
  S.~A.~Abel, C.~S.~Chu, J.~Jaeckel and V.~V.~Khoze,
  ``SUSY breaking by a metastable ground state: Why the early universe
  preferred the non-supersymmetric vacuum,''
  arXiv:hep-th/0610334.
}
\lref\CraigKX{
  N.~J.~Craig, P.~J.~Fox and J.~G.~Wacker,
  ``Reheating metastable O'Raifeartaigh models,''
  arXiv:hep-th/0611006.
}
\lref\FischlerXH{
  W.~Fischler, V.~Kaplunovsky, C.~Krishnan, L.~Mannelli and M.~Torres,
  ``Meta-stable supersymmetry breaking in a cooling universe,''
  arXiv:hep-th/0611018.
}
\lref\AbelMY{
  S.~A.~Abel, J.~Jaeckel and V.~V.~Khoze,
  ``Why the early universe preferred the non-supersymmetric vacuum. II,''
  arXiv:hep-th/0611130.
}
\lref\ShifmanUA{
  M.~A.~Shifman,
  ``Nonperturbative dynamics in supersymmetric gauge theories,''
  Prog.\ Part.\ Nucl.\ Phys.\  {\bf 39}, 1 (1997)
  [arXiv:hep-th/9704114].
}
\lref\EllisVI{
  J.~R.~Ellis, C.~H.~Llewellyn Smith and G.~G.~Ross,
  ``Will The Universe Become Supersymmetric?,''
  Phys.\ Lett.\ B {\bf 114}, 227 (1982).
}
\lref\DavisMZ{
  A.~C.~Davis, M.~Dine and N.~Seiberg,
  ``The Massless Limit Of Supersymmetric QCD,''
  Phys.\ Lett.\ B {\bf 125}, 487 (1983).
}

\lref\WittenDF{
  E.~Witten,
  ``Constraints On Supersymmetry Breaking,''
  Nucl.\ Phys.\ B {\bf 202}, 253 (1982).
}

\lref\NelsonNF{
  A.~E.~Nelson and N.~Seiberg,
  ``R symmetry breaking versus supersymmetry breaking,''
  Nucl.\ Phys.\ B {\bf 416}, 46 (1994)
  [arXiv:hep-ph/9309299].
}
\lref\RayWK{
  S.~Ray,
  ``Some properties of meta-stable supersymmetry-breaking vacua in Wess-Zumino
  models,''
  Phys.\ Lett.\ B {\bf 642}, 137 (2006)
  [arXiv:hep-th/0607172].
}
\lref\NSd{
  N.~Seiberg,
  ``Electric - magnetic duality in supersymmetric non-Abelian gauge theories,''
  Nucl.\ Phys.\ B {\bf 435}, 129 (1995)
  [arXiv:hep-th/9411149].
}
\lref\ISrev{
  K.~A.~Intriligator and N.~Seiberg,
  ``Lectures on supersymmetric gauge theories and electric-magnetic  duality,''
  Nucl.\ Phys.\ Proc.\ Suppl.\  {\bf 45BC}, 1 (1996)
  [arXiv:hep-th/9509066].
}
\lref\SeibergBZ{
  N.~Seiberg,
  ``Exact results on the space of vacua of four-dimensional SUSY gauge
  theories,''
  Phys.\ Rev.\ D {\bf 49}, 6857 (1994)
  [arXiv:hep-th/9402044].
}

\lref\FayetJB{
  P.~Fayet and J.~Iliopoulos,
 ``Spontaneously broken supergauge symmetries and Goldstone
  spinors,''
  Phys.\ Lett.\ B {\bf 51}, 461 (1974).
}

\lref\BaggerHH{
  J.~Bagger, E.~Poppitz and L.~Randall,
  ``The R axion from dynamical supersymmetry breaking,''
  Nucl.\ Phys.\ B {\bf 426}, 3 (1994)
  [arXiv:hep-ph/9405345].
}
\lref\IntriligatorRX{
  K.~A.~Intriligator, N.~Seiberg and S.~H.~Shenker,
  ``Proposal for a simple model of dynamical SUSY breaking,''
  Phys.\ Lett.\ B {\bf 342}, 152 (1995)
  [arXiv:hep-ph/9410203].
}

\lref\ISS{
  K.~Intriligator, N.~Seiberg and D.~Shih,
  ``Dynamical SUSY breaking in meta-stable vacua,''
  JHEP {\bf 0604}, 021 (2006)
  [arXiv:hep-th/0602239].
}

\lref\DineYW{
  M.~Dine and A.~E.~Nelson,
  ``Dynamical supersymmetry breaking at low-energies,''
  Phys.\ Rev.\ D {\bf 48}, 1277 (1993)
  [arXiv:hep-ph/9303230].
}
\lref\ISSn{
 K.~Intriligator, N.~Seiberg and D.~Shih,
  ``Supersymmetry Breaking, R-Symmetry Breaking and Metastable Vacua,''
  arXiv:hep-th/0703281.
}

\lref\IT{
  K.~Intriligator and S.~D.~Thomas,
  ``Dynamical Supersymmetry Breaking on Quantum Moduli Spaces,''
  Nucl.\ Phys.\ B {\bf 473}, 121 (1996)
  [arXiv:hep-th/9603158].
}
\lref\IY{
  K.~I.~Izawa and T.~Yanagida,
  ``Dynamical Supersymmetry Breaking in Vector-like Gauge Theories,''
  Prog.\ Theor.\ Phys.\  {\bf 95}, 829 (1996)
  [arXiv:hep-th/9602180].
}
\lref\ORaifeartaighPR{
  L.~O'Raifeartaigh,
  ``Spontaneous Symmetry Breaking For Chiral Scalar Superfields,''
  Nucl.\ Phys.\ B {\bf 96}, 331 (1975).
}
\lref\DvaliXE{
  G.~R.~Dvali and M.~A.~Shifman,
  ``Domain walls in strongly coupled theories,''
  Phys.\ Lett.\ B {\bf 396}, 64 (1997)
  [Erratum-ibid.\ B {\bf 407}, 452 (1997)]
  [arXiv:hep-th/9612128].
}
\lref\AffleckUZ{
  I.~Affleck, M.~Dine and N.~Seiberg,
  ``Calculable Nonperturbative Supersymmetry Breaking,''
  Phys.\ Rev.\ Lett.\  {\bf 52}, 1677 (1984).
}

\lref\FrancoES{
  S.~Franco and A.~M.~Uranga,
  ``Dynamical SUSY breaking at meta-stable minima from D-branes at obstructed
  geometries,''
  JHEP {\bf 0606}, 031 (2006)
  [arXiv:hep-th/0604136].
}

\lref\KitanoWM{
  R.~Kitano,
  ``Dynamical GUT breaking and mu-term driven supersymmetry breaking,''
  arXiv:hep-ph/0606129.
}

\lref\OoguriPJ{
  H.~Ooguri and Y.~Ookouchi,
  ``Landscape of supersymmetry breaking vacua in geometrically realized gauge
  theories,''
  Nucl.\ Phys.\ B {\bf 755}, 239 (2006)
  [arXiv:hep-th/0606061].
}

\lref\KitanoWZ{
  R.~Kitano,
  ``Gravitational gauge mediation,''
  Phys.\ Lett.\ B {\bf 641}, 203 (2006)
  [arXiv:hep-ph/0607090].
}

\lref\AmaritiVK{
  A.~Amariti, L.~Girardello and A.~Mariotti,
  ``Non-supersymmetric meta-stable vacua in SU(N) SQCD with adjoint matter,''
  arXiv:hep-th/0608063.
}
\lref\ShadmiJY{
  Y.~Shadmi and Y.~Shirman,
  ``Dynamical supersymmetry breaking,''
  Rev.\ Mod.\ Phys.\  {\bf 72}, 25 (2000)
  [arXiv:hep-th/9907225].
}
\lref\DineGM{
  M.~Dine, J.~L.~Feng and E.~Silverstein,
  ``Retrofitting O'Raifeartaigh models with dynamical scales,''
  Phys.\ Rev.\ D {\bf 74}, 095012 (2006)
  [arXiv:hep-th/0608159].
}
\lref\IntriligatorUK{
  K.~A.~Intriligator,
  ``'Integrating in' and exact superpotentials in 4-d,''
  Phys.\ Lett.\ B {\bf 336}, 409 (1994)
  [arXiv:hep-th/9407106].
}
\lref\SeibergBZ{
  N.~Seiberg,
  ``Exact Results On The Space Of Vacua Of Four-Dimensional Susy Gauge
  Theories,''
  Phys.\ Rev.\ D {\bf 49}, 6857 (1994)
  [arXiv:hep-th/9402044].
}
\lref\FinnellDR{
  D.~Finnell and P.~Pouliot,
  ``Instanton Calculations Versus Exact Results In Four-Dimensional Susy Gauge
  Theories,''
  Nucl.\ Phys.\ B {\bf 453}, 225 (1995)
  [arXiv:hep-th/9503115].
}
\lref\AharonyMY{
 O.~Aharony and N.~Seiberg,
 ``Naturalized and simplified gauge mediation,''
 arXiv:hep-ph/0612308.
}
\lref\NovikovPX{
  V.~A.~Novikov, M.~A.~Shifman, A.~I.~Vainshtein and V.~I.~Zakharov,
  ``Instantons In Supersymmetric Theories,''
  Nucl.\ Phys.\ B {\bf 223}, 445 (1983).
}
\lref\IntriligatorJR{
  K.~A.~Intriligator, R.~G.~Leigh and N.~Seiberg,
  ``Exact superpotentials in four-dimensions,''
  Phys.\ Rev.\ D {\bf 50}, 1092 (1994)
  [arXiv:hep-th/9403198].
}
\lref\WittenNF{
  E.~Witten,
  ``Dynamical Breaking Of Supersymmetry,''
  Nucl.\ Phys.\ B {\bf 188}, 513 (1981).
}
\lref\ArkaniHamedUT{
  N.~Arkani-Hamed and H.~Murayama,
  ``Renormalization group invariance of exact results in supersymmetric  gauge
  theories,''
  Phys.\ Rev.\ D {\bf 57}, 6638 (1998)
  [arXiv:hep-th/9705189].
}
\lref\DineXT{
  M.~Dine and J.~Mason,
  ``Gauge mediation in metastable vacua,''
  arXiv:hep-ph/0611312.
}

\lref\KitanoXG{
  R.~Kitano, H.~Ooguri and Y.~Ookouchi,
  ``Direct mediation of meta-stable supersymmetry breaking,''
  arXiv:hep-ph/0612139.
}
\lref\EtoYV{
  M.~Eto, K.~Hashimoto and S.~Terashima,
  ``Solitons in supersymmety breaking meta-stable vacua,''
  arXiv:hep-th/0610042.
}
\lref\MurayamaYF{
  H.~Murayama and Y.~Nomura,
  ``Gauge Mediation Simplified,''
  arXiv:hep-ph/0612186.
}
\lref\CsakiWI{
  C.~Csaki, Y.~Shirman and J.~Terning,
  ``A Simple Model of Low-scale Direct Gauge Mediation,''
  arXiv:hep-ph/0612241.
}
\lref\TerningTH{
  J.~Terning,
  ``Non-perturbative supersymmetry,''
  arXiv:hep-th/0306119.
}
\lref\MurayamaPB{
  H.~Murayama,
  ``A model of direct gauge mediation,''
  Phys.\ Rev.\ Lett.\  {\bf 79}, 18 (1997)
  [arXiv:hep-ph/9705271].
}
\lref\SeibergVC{
  N.~Seiberg,
  ``Naturalness versus supersymmetric nonrenormalization theorems,''
  Phys.\ Lett.\ B {\bf 318}, 469 (1993)
  [arXiv:hep-ph/9309335].
}
\lref\DimopoulosWW{
  S.~Dimopoulos, G.~R.~Dvali, R.~Rattazzi and G.~F.~Giudice,
  ``Dynamical soft terms with unbroken supersymmetry,''
  Nucl.\ Phys.\ B {\bf 510}, 12 (1998)
  [arXiv:hep-ph/9705307].
}
\lref\LutyNY{
  M.~A.~Luty,
  ``Simple gauge-mediated models with local minima,''
  Phys.\ Lett.\ B {\bf 414}, 71 (1997)
  [arXiv:hep-ph/9706554].
}

\lref\DimopoulosJE{
  S.~Dimopoulos, G.~R.~Dvali and R.~Rattazzi,
  ``A simple complete model of gauge-mediated SUSY-breaking and dynamical
  relaxation mechanism for solving the mu problem,''
  Phys.\ Lett.\ B {\bf 413}, 336 (1997)
  [arXiv:hep-ph/9707537].
}

\lref\ColemanPY{
  S.~R.~Coleman,
  ``The Fate Of The False Vacuum. 1. Semiclassical Theory,''
  Phys.\ Rev.\ D {\bf 15}, 2929 (1977)
  [Erratum-ibid.\ D {\bf 16}, 1248 (1977)].
}
\lref\AffleckXZ{
  I.~Affleck, M.~Dine and N.~Seiberg,
  ``Dynamical Supersymmetry Breaking In Four-Dimensions And Its
  Phenomenological Implications,''
  Nucl.\ Phys.\ B {\bf 256}, 557 (1985).
}

\lref\AffleckVC{
  I.~Affleck, M.~Dine and N.~Seiberg,
  ``Dynamical Supersymmetry Breaking In Chiral Theories,''
  Phys.\ Lett.\ B {\bf 137}, 187 (1984).
}

\lref\WittenNF{
  E.~Witten,
  ``Dynamical Breaking Of Supersymmetry,''
  Nucl.\ Phys.\ B {\bf 188}, 513 (1981).
}

\lref\MeuriceAI{
  Y.~Meurice and G.~Veneziano,
  ``Susy Vacua Versus Chiral Fermions,''
  Phys.\ Lett.\ B {\bf 141}, 69 (1984).
}

\lref\AffleckMK{
  I.~Affleck, M.~Dine and N.~Seiberg,
  ``Dynamical Supersymmetry Breaking In Supersymmetric QCD,''
  Nucl.\ Phys.\ B {\bf 241}, 493 (1984).
}

\lref\AffleckUZ{
  I.~Affleck, M.~Dine and N.~Seiberg,
  ``Calculable Nonperturbative Supersymmetry Breaking,''
  Phys.\ Rev.\ Lett.\  {\bf 52}, 1677 (1984).
}

\lref\LutyNQ{
  M.~A.~Luty and J.~Terning,
  ``New mechanisms of dynamical supersymmetry breaking and direct gauge
  mediation,''
  Phys.\ Rev.\ D {\bf 57}, 6799 (1998)
  [arXiv:hep-ph/9709306].
}

\lref\CsakiGR{
  C.~Csaki, L.~Randall, W.~Skiba and R.~G.~Leigh,
  ``Supersymmetry breaking through confining and dual theory gauge  dynamics,''
  Phys.\ Lett.\ B {\bf 387}, 791 (1996)
  [arXiv:hep-th/9607021].
}

\lref\PoppitzWP{
  E.~Poppitz, Y.~Shadmi and S.~P.~Trivedi,
  ``Supersymmetry breaking and duality in SU(N) x SU(N-M) theories,''
  Phys.\ Lett.\ B {\bf 388}, 561 (1996)
  [arXiv:hep-th/9606184].
}
\lref\MurayamaYF{
  H.~Murayama and Y.~Nomura,
  ``Gauge Mediation Simplified,''
  arXiv:hep-ph/0612186.
}
\lref\CsakiWI{
  C.~Csaki, Y.~Shirman and J.~Terning,
  ``A Simple Model of Low-scale Direct Gauge Mediation,''
  arXiv:hep-ph/0612241.
}

\lref\MurayamaPB{
  H.~Murayama,
  ``A model of direct gauge mediation,''
  Phys.\ Rev.\ Lett.\  {\bf 79}, 18 (1997)
  [arXiv:hep-ph/9705271].
}

\lref\DimopoulosWW{
  S.~Dimopoulos, G.~R.~Dvali, R.~Rattazzi and G.~F.~Giudice,
  ``Dynamical soft terms with unbroken supersymmetry,''
  Nucl.\ Phys.\ B {\bf 510}, 12 (1998)
  [arXiv:hep-ph/9705307].
}

\lref\LutyNY{
  M.~A.~Luty,
  ``Simple gauge-mediated models with local minima,''
  Phys.\ Lett.\ B {\bf 414}, 71 (1997)
  [arXiv:hep-ph/9706554].
}

\lref\DimopoulosJE{
  S.~Dimopoulos, G.~R.~Dvali and R.~Rattazzi,
  ``A simple complete model of gauge-mediated SUSY-breaking and dynamical
  relaxation mechanism for solving the mu problem,''
  Phys.\ Lett.\ B {\bf 413}, 336 (1997)
  [arXiv:hep-ph/9707537].
}

\lref\VenezianoAH{
  G.~Veneziano and S.~Yankielowicz,
  ``An Effective Lagrangian For The Pure N=1 Supersymmetric Yang-Mills
  Theory,''
  Phys.\ Lett.\ B {\bf 113}, 231 (1982).
}

\lref\TaylorBP{
  T.~R.~Taylor, G.~Veneziano and S.~Yankielowicz,
  ``Supersymmetric QCD And Its Massless Limit: An Effective Lagrangian
  Analysis,''
  Nucl.\ Phys.\ B {\bf 218}, 493 (1983).
}

\lref\AffleckVC{
  I.~Affleck, M.~Dine and N.~Seiberg,
  ``Dynamical Supersymmetry Breaking In Chiral Theories,''
  Phys.\ Lett.\ B {\bf 137}, 187 (1984).
}
\lref\WessCP{
  J.~Wess and J.~Bagger,
  ``Supersymmetry and supergravity.''
}

\lref\WeinbergCR{
  S.~Weinberg,
  ``The quantum theory of fields.  Vol. 3: Supersymmetry.''
}

\lref\HuqUE{
  M.~Huq,
  ``On Spontaneous Breakdown Of Fermion Number Conservation And
  Supersymmetry,''
  Phys.\ Rev.\ D {\bf 14}, 3548 (1976).
}

\lref\TerningBQ{
  J.~Terning,
  ``Modern supersymmetry: Dynamics and duality.''
}

\lref\Dinebook{M.~Dine, ``Supersymmetry and String Theory: Beyond
the Standard Model.''}

\lref\FerraraQS{
  S.~Ferrara, L.~Girardello and H.~P.~Nilles,
  ``Breakdown Of Local Supersymmetry Through Gauge Fermion Condensates,''
  Phys.\ Lett.\ B {\bf 125}, 457 (1983).
}

\lref\DerendingerKK{
  J.~P.~Derendinger, L.~E.~Ibanez and H.~P.~Nilles,
  ``On The Low-Energy D = 4, N=1 Supergravity Theory Extracted From The D = 10,
  N=1 Superstring,''
  Phys.\ Lett.\ B {\bf 155}, 65 (1985).
}

\lref\DineRZ{
  M.~Dine, R.~Rohm, N.~Seiberg and E.~Witten,
  ``Gluino Condensation In Superstring Models,''
  Phys.\ Lett.\ B {\bf 156}, 55 (1985).
}

\Title{\vbox{\baselineskip12pt \hbox{UCSD-PTH-07-02}}}
{\vbox{\centerline{Lectures on Supersymmetry Breaking}}}
\smallskip
\centerline{Kenneth Intriligator$^{1}$ and Nathan Seiberg$^2$}
\smallskip
\bigskip
\centerline{$^1${\it Department of Physics, University of
California, San Diego, La Jolla, CA 92093 USA}}
\medskip
\centerline{$^2${\it School of Natural Sciences, Institute for
Advanced Study, Princeton, NJ 08540 USA}}
\bigskip
\vskip 1cm

 \noindent
We review the subject of spontaneous supersymmetry breaking. First
we consider supersymmetry breaking in a semiclassical theory. We
illustrate it with several examples, demonstrating different
phenomena, including metastable supersymmetry breaking.  Then we
give a brief review of the dynamics of supersymmetric gauge
theories. Finally, we use this dynamics to present various
mechanisms for dynamical supersymmetry breaking.  These notes are
based on lectures given by the authors in 2007, at various
schools.

\bigskip

 \Date{February 2007}

\newsec{Introduction}

With the advent of the LHC it is time to review old model building
issues leading to phenomena which could be discovered, or
disproved, by the LHC. Supersymmetry (SUSY) is widely considered
as the most compelling new physics that the LHC could discover. It
gives a solution to the hierarchy problem, leads to coupling
constant unification and has dark matter candidates.

Clearly, the standard model particles are not degenerate with
their superpartners, and therefore supersymmetry should be broken.
To preserve the appealing features of supersymmetry, this breaking
must be spontaneous, rather than explicit breaking.
This means that the Lagrangian is supersymmetric, but the vacuum
state is not invariant under supersymmetry.

Furthermore, as was first suggested by Witten \WittenNF, we would
like the mechanism which spontaneously breaks supersymmetry to be
dynamical.  This means that it arises from an exponentially small
effect, and therefore it naturally leads to a scale of
supersymmetry breaking, $M_s$, which is much smaller than the high
energy scales in the problem $M_{cutoff}$ (which can be the Planck
scale or the grand unified scale):
 \eqn\scaleofs{M_s = M_{cutoff}e^{-c/g(M_{cutoff})^2}\ll
 M_{cutoff}.}
This can naturally lead to hierarchies.  For example, the weak
scale $m_W$ can be dynamically generated, explaining why
$m_W/m_{Pl}\sim 10 ^{-17}$.

In these lectures, we will focus on the key conceptual issues and
mechanisms for supersymmetry breaking, illustrating them with the
simplest examples. We will not discuss more detailed model
building questions, such as the question of how the supersymmetry
breaking is mediated to the MSSM, and what the experimental
signatures of the various mediation schemes are. These are very
important topics, which deserve separate sets of lectures.  Also,
we will not discuss supersymmetry breaking by Fayet-Iliopoulos
terms \FayetJB.

We will assume that the readers (and audience in the lectures)
have some basic familiarity with supersymmetry.  Good textbooks
are \refs{\WessCP\GatesNR\WeinbergCR\TerningBQ-\Dinebook}.

As seen from the supersymmetry algebra,
 \eqn\susyal{\{Q_\alpha, \overline Q_{\dot \alpha}\}=
 2P_{\alpha \dot \alpha},}
the vacuum energy
 \eqn\vacen{ \langle \psi |{\cal H}|\psi
 \rangle \propto \sum _\alpha \big| Q_\alpha |\psi \rangle \big
 |^2+\sum _{\dot \alpha}\big| \overline Q_{\dot \alpha} |\psi
 \rangle \big |^2 \geq 0}
is an order parameter for supersymmetry breaking.  Supersymmetry
is spontaneously broken if and only if the vacuum has non-zero
energy\foot{In these lectures we focus on global SUSY,
$M_{pl}\rightarrow \infty$.  In supergravity we can add an
arbitrary negative constant to the vacuum energy, via $\Delta
W=const$, so the cosmological constant can still be tuned to the
observed value.},
 \eqn\vaceng{V_{vac}=M_s^4. }
In the case of dynamical supersymmetry breaking (DSB), the scale
$M_s$ is generated by dimensional transmutation, as in \scaleofs.

As with the spontaneous breaking of an ordinary global symmetry,
the broken supersymmetry charge $Q$ does not exist in an infinite
volume system. Instead, the supersymmetry current $S$ exists, and
its action on the vacuum creates a massless particle -- the
Goldstino.  (The supercharge tries to create a zero momentum
Goldstino, which is not normalizable.)  In the case of
supergravity, where the symmetry \susyal\ is gauged, we have the
standard Higgs mechanism and the massless Goldstino is ``eaten''
by the gravitino.

There are many challenges in trying to implement realistic
realizations of dynamical supersymmetry breaking.  A first
challenge, which follows from the Witten index \WittenDF, is that
dynamical supersymmetry breaking, where the true vacuum is static
and has broken supersymmetry, seems non-generic, requiring
complicated looking theories.  On the other hand, accepting the
possibility  that we live in a metastable vacuum improves the
situation.  As even very simple theories can exhibit metastable
dynamical supersymmetry breaking, it could be generic \ISS.
(Particular models of metastable supersymmetry breaking have been
considered long ago, e.g.\ a model \EllisVI, which we review
below.)

Another challenge is the relation \NelsonNF\ between  R-symmetry
and broken supersymmetry.  Generically, there is broken
supersymmetry if and only if there is an R-symmetry.  As we will
also discuss, there is broken supersymmetry in a metastable state
if and only if there is an approximate R-symmetry.  For building
realistic models, an unbroken R-symmetry is problematic.  It
forbids Majorana gaugino masses.  Having an exact, but
spontaneously broken R-symmetry is also problematic, it leads to a
light R-axion (though including gravity can help\foot{Including
gravity, the R-symmetry needs to be explicitly broken, in any
case, by the  $\Delta W={\rm const.}$, needed  to get a realistic
cosmological constant.  It is possible that this makes the R-axion
sufficiently massive \BaggerHH.}). We are thus led to explicitly
break the R-symmetry.  Ignoring gravity, this then means that we
should live in a metastable state!

The outline of these lectures is as follows.  In the next section,
we consider theories in which the supersymmetry breaking can be
seen semiclassically.   Such theories can arise as the low energy
theory of another microscopic theory.  Various general points about
supersymmetry breaking (or restoration) are illustrated, via
several simple examples.

In section 3, we give a lightning review of $\CN =1$
supersymmetric QCD (SQCD), with various numbers of colors and
flavors. Here we will be particularly brief.  The reader can
consult various books and reviews, e.g.\
\refs{\TerningBQ,\Dinebook,\ISrev\PeskinQI\ShifmanUA-\StrasslerQG},
for more details.

In section 4, we discuss dynamical supersymmetry breaking (DSB),
where the supersymmetry breaking is related to a dynamical scale
$\Lambda$, and thus it is non-perturbative in the coupling.  Using
the understood dynamics of SQCD, it is possible to find an
effective Lagrangian in which supersymmetry breaking can be seen
semiclassically. We will discuss only four characteristic
examples, demonstrating four different mechanisms of DSB.

\newsec{Semiclassical spontaneous supersymmetry breaking}

In this section we consider theories with chiral superfields
$\Phi^a$, a smooth K\"ahler potential $K(\Phi, \bar \Phi)$ and a
superpotential $W(\Phi)$.  For simplicity we will ignore the
possibility of adding gauge fields.  A detailed analysis of their
effect will be presented in \ISSn.  The K\"ahler potential leads
to the metric on field space
 \eqn\metricK{g_{a \bar a} = \partial_a \partial_{\bar a} K,}
which determines the Lagrangian of the scalars
 \eqn\scalarL{\eqalign{
 \CL_{scalars}= & g_{a\bar a}\partial_\mu \Phi^a \partial^\mu \bar
 \Phi^{\bar a} - V(\Phi,\bar \Phi)\cr
 V=& g^{a\bar a}\partial_a W \partial_{\bar a} \bar W.}}
It is clear from the scalar potential $V$ that supersymmetric
ground states, which must have zero energy, are related to the
critical points of $W$; i.e.\ points where we can solve
 \eqn\criticalW{\partial_a W(\Phi^a) =0 \qquad \forall a .}
If no such point exists, it means that the system does not have
supersymmetric ground states.

However, before we conclude in this case that supersymmetry is
spontaneously broken we should also exclude the possibility that
the potential slopes to zero at infinity.  Roughly, in this case
the system has ``a supersymmetric state at infinity.''  More
precisely, it does not have a ground state at all!

\subsec{The simplest example}

Consider  a theory of a single chiral superfield $X$, with linear
superpotential with coefficient $f$ (with units of mass square),
 \eqn\wlin{W =f X,}
and canonical K\"ahler potential
 \eqn\cansfK{K=K_{can}=\bar XX.}
Supersymmetry is spontaneously broken by the expectation value of
the F-component of $X$, $\bar F_X = -f$.  Using \scalarL\ the
potential is $V=|f|^2$.  It is independent of $X$, so there are
classical vacua for any $\ev{X}$.

Supersymmetric theories often have a continuous manifold of
supersymmetric vacua which are usually referred to as ``moduli
space of vacua.'' However, in the case where supersymmetry is
broken, such a space is not robust: this nonsupersymmetric
degeneracy of vacua is often lifted once radiative corrections are
taken into account.  Therefore, we prefer to refer to this space
as a {\it pseudomoduli space of vacua}.  The example we
study here is free, and therefore the space of vacua remains
present even in the quantum theory.  We will see below examples of
the more typical situation, in which the classical theory has a
pseudomoduli space of nonsupersymmetric vacua, but the quantum
corrections lift the degeneracy.

The exactly massless Goldstino is $\psi _X$, and its complex
scalar partner $X$ is the classically massless pseudomodulus. Note
that there is a $U(1)_R$ symmetry, with $R(X)=2$. For $\ev{X}\neq
0$ it is spontaneously broken, and the corresponding massless
Goldstone boson is the phase of the field $X$.

Deforming \wlin\ by any superpotential interactions, say a degree
$n$ polynomial in $X$, leads to $n-1$ supersymmetric vacua. For
example, if we add $\Delta W=\half \epsilon X^2$, there is a
vacuum with unbroken supersymmetry at $\ev{X}=-f/\epsilon$. This
deformation lifts the pseudomoduli space by creating a potential
$|f + \epsilon X|^2$ over it.  We can also see that supersymmetry
is not broken from the fact that $\psi _X$ now has mass
$\epsilon$, so there is no massless Goldstino.  Note also that any
such $\Delta W$ deformations of \wlin\ explicitly break the
$U(1)_R$ symmetry; the fact that they lead to supersymmetric vacua
illustrates a general connection between R-symmetry and
supersymmetry breaking, which will be developed further below.

\subsec{The simplest example but with more general K\"ahler
potential}

Consider again the theory of section 2.1 with superpotential
\wlin, but with a general K\"ahler potential $K(X, \bar X)$.  Of
course, this theory is not renormalizable.  It should be viewed
either as a classical field theory or as a quantum field theory
with a cutoff $\Lambda$. More physically, such a theory can be the
low energy approximation of another, microscopic theory, which is
valid at energies larger than $\Lambda$.

The potential,
 \eqn\alsime{V=K_{X\bar X}^{-1}|f |^2}
lifts the degeneracy along the pseudomoduli space of the previous
example.  Let us suppose that the K\"ahler potential $K$ is
smooth. (Non-smooth $K$ signals the need to include additional
degrees of freedom, in the low-energy effective field theory at
the singularity.  An example of this case is discussed in the next
subsection.)  For smooth $K$, the potential \alsime\ is
non-vanishing, and thus there is no supersymmetric vacuum.

Before concluding that supersymmetry is spontaneously broken, we
should consider the behavior at $|X|\to \infty$. If there is any
direction along which $\lim_{|X| \to \infty} K_{X\bar X} $
diverges, then $V$ slopes to zero at infinity and the system does
not have a ground state. If $\lim_{|X| \to \infty} K_{X\bar X} $
vanishes in all directions, the potential rises  at infinity and
it has a supersymmetry breaking global minimum for some finite
$X$. Finally, if there are directions along which $\lim_{|X| \to
\infty} K_{X\bar X} $ is finite, the potential approaches a
constant along these directions and the global minimum of the
potential needs a more detailed analysis.

Consider the behavior of the system near a particular point, say
$X\approx 0$.  Let
 \eqn\apprK{K=X\bar X -{c\over |\Lambda |^2}(X\bar X)^2+\dots,}
with positive $c$.\foot{The parameter $\Lambda$ in \apprK\
determines the scale of the features in the potential.  When this
theory arises as the low energy approximation of another theory,
this parameter $\Lambda$ is typically the scale above which the
more microscopic theory is valid.}  Then there is a locally stable
nonsupersymmetric vacuum at $X=0$. In this vacuum, the scalar
component of $X$ gets mass $m_X^2= 4c|f |^2/|\Lambda |^2$.  The
fermion $\psi _X$ is the exactly massless Goldstino.  Note also
that if $K(X, \bar X)$ depends only on $X\bar X$, then there is a
$U(1)_R$ symmetry, which is unbroken if the vacuum is at $X=0$.
This ground state can be the global minimum of the potential.
Alternatively, it can be only a local minimum, with either another
minimum of lower energy or no minimum at all if the system runs
away to infinity.

If $X=0$ is not the global minimum of the potential, the state at
$X=0$ is metastable.  If the theory is sufficiently weakly
coupled, the tunneling out of this vacuum can be highly suppressed
and this vacuum can be very long lived. We see that it is easy to
find examples where supersymmetry is broken in a long lived
metastable state. (Though we have not yet demonstrated what
physical dynamics leads to such features in the K\"ahler
potential.)

Let us consider again the theory with K\"ahler potential \apprK,
but deform the superpotential \wlin\ to
 \eqn\wdefm{W=fX+\half \epsilon X^2,}
taking $\epsilon$ as a small parameter.  There is now a supersymmetric vacuum at
\eqn\Xsusy{\ev{X}_{susy}=-f/\epsilon,} which is very far from the
origin.  On the other hand, for $X$ near the origin, we find for
the potential
 \eqn\vxnz{V(X, \bar X)=(K_{X\bar
 X})^{-1}|f+\epsilon X|^2= |f|^2 +\bar f\epsilon X +f\bar \epsilon
 \bar X
 +{4c|f|^2\over |\Lambda |^2}|X|^2+\dots  \qquad (X\approx
 0, \ \epsilon \ll 1).}
There is a local minimum, with broken supersymmetry, at
 \eqn\xlocmin{\ev{X}_{meta}=-{\bar  \epsilon  |\Lambda |^2\over
 4c \bar f}.}
For $|\epsilon |\ll \sqrt{c}|f/\Lambda|$, this supersymmetry breaking vacuum is very
far from the supersymmetric vacuum \Xsusy. The metastable state
\xlocmin\ can thus be very long lived.

At first glance, there is a small puzzle with the broken
supersymmetry vacuum \xlocmin.  The superpotential \wdefm\ gives a
mass $\epsilon$ to the fermion $\psi _X$, whereas any vacuum with
broken supersymmetry must have an {\it exactly massless}
Goldstino.  The Goldstino must be exactly massless, regardless of
whether the supersymmetry breaking state is a local or global
minimum of the potential. The resolution of the
apparent puzzle is that
 \eqn\kcontains{\int d^4 \theta K\supset K_{XX\bar X}\bar F_X
 \psi _X \psi _X}
and evaluating this term in the vacuum \xlocmin, with $\bar F_X\approx
-f$, exactly cancels the $\epsilon \psi \psi$ term coming from
the superpotential.  So there is indeed an exactly massless
Goldstino, $\psi _X$, consistent with the supersymmetry breaking
in the metastable state.

\subsec{Additional degrees of freedom can restore supersymmetry}

Let us consider a renormalizable theory of two chiral superfields,
$X$ and $q$, with canonical K\"ahler potential, $K=X\bar X+q\bar
q$.  We modify the example of section 2.1 by coupling the field
$X$ to the additional field $q$ via
 \eqn\wlinq{W=\half hXq^2+fX,}
where $h$ is the coupling constant.  The field $q$ gets a mass
from an $X$ expectation value (an added mass term $\Delta W=\half
M q^2$ can be eliminated by a shift of $X$). There is a $U(1)_R$
symmetry, with $R(X)=2$, and $R(q)=0$, and also a ${\bf Z}_2$
symmetry $q\to -q$.

The potential
 \eqn\pottt{ V=|h X q|^2 + |\half h q^2 + f|^2}
does not break supersymmetry.  There are two supersymmetric vacua,
at
 \eqn\simpexsusy{\ev{X}_{susy}=0, \qquad \ev{q}_{susy}=\pm
 \sqrt{-2f/h}.}
The additional degrees of freedom, $q$, as compared with the
example of section 2.1, have restored supersymmetry.

Note that the potential \pottt\ also has a supersymmetry breaking
pseudoflat direction with $\ev{q}=0 $, and arbitrary $\ev{X}$,
with $V=|f|^2$.  It reflects the fact that for large $X$ the $q$
fields are massive, can be integrated out, and the low energy
theory is then the same as that of section 2.1.  The spectrum of
the massive $q$ fields depends on $X$, and is given by
 \eqn\masspss{m_0^2= |hX|^2 \pm |hf| \qquad ; \qquad m_{1/2} =
 hX.}
We see, however, that this pseudomoduli space has a tachyon for
 \eqn\notatt{|X|^2<  \left|{f \over  h}\right|.}
In the region \notatt, the potential can decrease along the
$\ev{q}$ direction,  down to the supersymmetric vacua \simpexsusy.

\subsec{An example with a runaway \WittenKV}

Consider a renormalizable theory of two chiral superfields, $X$
and $Y$, with canonical K\"ahler potential, and superpotential
 \eqn\worun{W=\half h X^2Y +fX.}
There is a $U(1)_R$ symmetry, with $R(X)=2$, and $R(Y)=-2$. The
potential is
 \eqn\vorun{V=\left| \half h X^2\right| ^2+\left|
 hXY+f\right| ^2.}
It is impossible for both terms to vanish, so the theory does not
have supersymmetric ground states.  As usual, before concluding
that supersymmetry is spontaneously broken, we must examine for
runaway directions. Indeed, taking $X=-f/hY$ the potential has a
runaway direction as $Y\to \infty$:
 \eqn\vorunn{V\to {\left| { f^2\over 2hY^2}\right| ^2}\to 0.}
There is no static vacuum, but supersymmetry is asymptotically
restored as $Y\to \infty$.

For large $|Y|$ the supersymmetry breaking is small, and the
mass of $X$ is large,  so  we can
describe the theory by a supersymmetric low-energy effective
Lagrangian with $X$ integrated out.  Integrating out $X$ in \worun\ we find the effective
superpotential
 \eqn\wrune{W_{eff}=-{f^2 \over 2h Y}}
which is consistent with the R-symmetry, and leads to the potential
\vorunn.

\subsec{O'Raifeartaigh-type models}

Here we discuss models of supersymmetry breaking which arise in
renormalizable field theories; i.e.\ unlike the example of section
2.2, we will examine classical theories with a canonical K\"ahler
potential (for a recent analysis of such models see e.g.\ \RayWK).

The simplest version of this class of models has three chiral
superfields, $X_1$, $X_2$, and $\phi$, with canonical K\"ahler
potential
 \eqn\ORaiK{K_{cl}=\bar X_1 X_1+\bar X _2 X_2+ \bar \phi  \phi}
and superpotential
 \eqn\worg{W=X_1g_1(\phi )+X_2g_2(\phi)}
with quadratic polynomials $g_{1,2}(\phi)$. This theory has a
$U(1)_R$ symmetry, with $R(X_1)=R(X_2)=2$, and $R(\phi )=0$.  The
tree-level potential for the scalars is
 \eqn\orpot{V_{tree}=\left|
 F_{X_1}\right| ^2 + \left| F_{X_2} \right| ^2+ \left| F_{\phi
 }\right|^2}
with
 \eqn\for{ -\bar F_{X_1}=\partial_{X_1} W= g_1(\phi), \quad -\bar
 F_{X_2}=g_2(\phi), \quad
 -\bar F_{\phi}=X_1 g'_1(\phi) +X_2g_2'(\phi).}
We are interested in the minima of this potential.

We can always choose $X_1$ and $X_2$ to set $ F_\phi=0$.  But, for
generic functions $g_1(\phi )$ and $g_2(\phi)$, we cannot
simultaneously solve $g_1(\phi) =0$ and $g_2(\phi )=0$, so
$F_{X_1}$ or $ F_{X_2}$ is non-zero, and hence supersymmetry is
generically broken.   There is a one-complex dimensional classical
pseudomoduli space of non-supersymmetric vacua, since only one
linear combination of $X_1$ and $X_2$ is constrained by the
condition that $ F_\phi =0$. Setting $F_\phi =0$ ensures that the
vacuum satisfies the $X_1$ and $X_2$ equations of motion,
$\partial _{X_i}V_{tree}=0$. We still need to impose $\partial
_\phi V_{tree}=0$, which requires that $\ev{\phi}$ solve
 \eqn\phieom{\overline{g_1(\phi )}
 g_1'(\phi)+\overline{g_2(\phi)}g_2'(\phi)=0.}

Expanding to quadratic order in $\delta X_1$, $\delta X_2$, and
$\delta \phi$ yields the mass matrix $m_0^2$ of the massive
scalars; the eigenvalues of this matrix must all be non-negative,
of course, if we are expanding around a (local) minimum of the
potential. The fermion mass terms are given by
 \eqn\fermilag{{\cal L}\supset
 (X_1g_1''(\phi )+X_2g_2''(\phi))\psi _\phi \psi _\phi
 +(g_1'(\phi)\psi _{X_1}+g_2'(\phi ) \psi _{X_2})\psi _\phi.}
It is easy to see that there is a massless eigenvector,
corresponding to the massless Goldstino.

\bigskip
\centerline{\it Example 1 -- the basic O'Raifeartaigh model
\ORaifeartaighPR}

As a special case of the above class of models, consider\foot{If,
instead, $g_{1,2}$ are even quadratic polynomials:
$g_i(\phi)=\half h_i \phi ^2+f_i$, a simple change of variables
shows that the theory decouples to a free field which breaks
supersymmetry as in section 2.1 and the example of section 2.3.}
$g_1(\phi)=\half h \phi ^2+f$, $g_2(\phi)=m\phi$.  It is
characterized by the discrete ${\bf Z}_2$ symmetry under which
$\phi$ and $X_2$ are odd.

For convenience, let us also write it as
 \eqn\wor{W=\half h X \phi _1^2+ m\phi _1\phi _2 +fX,}
where we denote $X=X_1$, $\phi _2=X_2$, and $\phi _1= \phi $. Note
that, for $m\to 0$, the field $\phi _2$ decouples, and what
remains in \wor\ is the theory of section 2.3, which we have seen
does not break supersymmetry.  For $m\neq 0$, it does break
supersymmetry, as in the general case discussed above,  as there
is no simultaneous solution of $g_1(\phi _1)=\half h\phi _1^2+f=0$
and $g_2(\phi _1 )=m\phi _1=0$.  The potential rises for large
$\phi _1$ and $\phi _2$, so these fields do not have runaway
directions. The minima of the potential form a one-complex
dimensional pseudomoduli space of degenerate, non-supersymmetric
vacua, with $\ev{X}$ arbitrary.

The equation \phieom\ is a cubic equation for $\phi_1$. The
solution with minimum energy depends on the parameter
 \eqn\ydef{y\equiv \left|{hf\over m^2}\right|.}
Consider the case $y<  1$.  Then the potential is
minimized\foot{There is a second order phase transition at $y=1$,
where this minimum splits to two minima and a saddle point.  Here
we will not analyze the phase $y>1$.  See e.g.\ \ISS\ for a
detailed analysis.} by $F_{\phi _2}=0$, with value
 \eqn\VminOR{V_{min}=|F_X|^2=|f|^2,}
at $\phi _1=\phi _2=0$ and arbitrary $X$.

The fermion $\psi _X$ is the exactly massless Goldstino.  The
scalar component of $X$ is a classical pseudomodulus. The
classical mass spectrum of the $\phi _1$ and $\phi _2$ fields can
be easily computed.  For the two, two-component fermions, the
eigenvalues are
 \eqn\eigenei{ m_{1/2}^2 = {1\over4} (|hX|\pm
 \sqrt{|hX|^2+4|m|^2})^2,}
and for the four real scalars the mass eigenvalues are
 \eqn\eigenii{ m_{0}^2 =\left(|m|^2+\half \eta |hf |
 +\half |hX|^2\pm\half \sqrt{|hf|^2+2\eta |hf|
 |hX|^2+4|m|^2|hX|^2+|hX|^4}\right), }
where $\eta =\pm 1$. We see that, as in \masspss, the spectrum
changes along the pseudomoduli space parameterized by $X$; these
vacua are physically distinct.

The parameter $y$ sets the relative size of the mass splittings,
corresponding to supersymmetry being broken, between \eigenei\ and
\eigenii. For $y\ll 1$, the spectrum \eigenei\ and \eigenii\ is
approximately supersymmetric, whereas for $y\sim 1$ supersymmetry
is badly broken. (In particular, for $y=1$, there is a massless
real scalar in \eigenii\  for all $X$, whereas the fermions
\eigenei\ are all massive.)

We can write \wor\ as $W=\half M_{ij}\phi ^i \phi ^j +fX$, where
$M=\pmatrix{hX& m\cr m&0}$, and the supersymmetry breaking can be
seen from the fact that $\det M=-m^2$ is non-zero and $X$
independent.  This can be generalized to similar models, with more
fields $\phi ^i$, and $M_{ij}$ such that $\det M$ is non-zero and
independent of $X$ \ISS.

\bigskip
\centerline{\it Example 2 -- supersymmetry breaking in a
metastable state \EllisVI}

We noted above that the theory \worg\ breaks supersymmetry for
generic functions $g_1(\phi)$ and $g_2(\phi)$, because we
generically cannot solve $g_1(\phi)=g_2(\phi)=0$. Let us consider
the case of a {\it non-generic} superpotential, where there is a
solution $\ev{\phi} _{susy}$ of $g_1(\phi )=g_2(\phi )=0$.  In
this case, there are supersymmetric vacua.  There can still,
however, be metastable vacua with broken supersymmetry.

As a particular example, consider
 \eqn\fiare{g_1(\phi )=h\phi (\phi -m_1), \qquad g_2(\phi )=
 m_2(\phi -m_1).}
(This theory was first analyzed in \EllisVI\ and was recently
reexamined in \RayWK.)  There is a moduli space of supersymmetric
vacua at
 \eqn\susyvame{\ev{\phi}_{susy}=m_1 \qquad ; \qquad \ev{X _2}_{susy}
 =-{h m_1\over m_2} \ev{X_1}_{susy},}
with arbitrary $\ev{X_1}_{susy}$.  The equation \phieom\ is a
cubic equation for $\phi$, and this moduli space of supersymmetric
vacua corresponds to one root of this cubic equation.  For
$|hm_1/m_2|^2>8$, there is also a pseudomoduli space of
supersymmetry violating minima of the potential at
 \eqn\nongex{\ev{\phi _1}_{meta}\approx \left| {m_2
 \over h m_1}\right| ^2 m_1\qquad , \qquad \ev{X _2}_{meta} \approx
 {h m_1\over m_2}\ev{X_1}_{meta}  \qquad \hbox{for}\ \  \left|
 {hm_1\over m_2}\right| \gg 1}
with arbitrary $\ev{X_1}_{meta}$. These metastable false vacua, in
which supersymmetry is broken, become parametrically long lived as
$|hm_1/m_2|$ is increased \EllisVI.  (The third root of the cubic
equation \phieom\ is a saddle point.)

\subsec{Metastable SUSY breaking in a modified O'Raifeartaigh
model \ISSn}

Let us modify the original, basic O'Raifeartaigh model by adding
to the superpotential \wor\ a small correction
 \eqn\worm{W=\half hX \phi _1^2+ m\phi _1\phi _2 +fX + \half
 \epsilon m\phi_2^2}
with $|\epsilon | \ll 1$.   This added term breaks the $U(1)_R$
symmetry.  It has an interesting effect: it leads to metastable supersymmetry breaking.
A similar model, but with the $\epsilon$ term in \worm\ replaced with $\half \epsilon m X^2$
was considered in \DineGM, with similar conclusions to ours here.  (Note that adding
$\Delta W=\half b\phi _1^2$ has no physical effect; it can simply
be eliminated by shifting $X$ by an appropriate constant.)

The potential is now
 \eqn\orpotm{V_{tree}=\left|
 F_X\right| ^2 + \left| F_{\phi _1} \right| ^2+ \left| F_{\phi
 _2}\right|^2}
with
 \eqn\form{ -\bar F_X= \half h\phi _1^2 +f,
 \quad -\bar F_{\phi _1}=
 hX\phi _1 +m\phi _2, \quad -\bar F_{\phi_2}=m\phi _1
 + \epsilon m \phi_2.}
Because of the modification of the superpotential by the last term
in \worm\ two new supersymmetric minima appear at
 \eqn\SUSYmo{\ev{\phi_1}_{susy}=\pm \sqrt{-2f/h} ,
 \qquad \ev{\phi_2}_{susy}=  \mp{1
 \over \epsilon} \sqrt{-2f/h}, \qquad \ev{X}_{susy}=
 { m \over h\epsilon} }
However, for small $\epsilon$ and $y=\left|{h f \over m^2} \right|
<1$, the potential near the previous supersymmetry breaking
minimum $\phi_1=\phi_2=0$ is not modified a lot.

Strictly, this theory does not break supersymmetry -- it has
supersymmetric ground states at \SUSYmo.  However, the
generalization of the eigenvalues \eigenii, to include $\epsilon$,
remains non-tachyonic for
 \eqn\nontapm{\left|X- {m\over h\epsilon}\right|^2  >
 \left( {1 \over |\epsilon|^2}+1 \right)\left|{f\over h}\right|.}
Therefore, most of the
pseudomoduli space of vacua of the $\epsilon=0$ theory remains
locally stable, and the tachyon exists only in a neighborhood of
the supersymmetric value \SUSYmo. In particular, for small
$\epsilon $ and $y<1$, the region near $X=0$ is locally stable.

As $\epsilon \to 0$ the supersymmetry preserving vacua \SUSYmo\
are pushed to infinity until finally, for $\epsilon=0$ they are
not present, and we are left with only the pseudomoduli space of
nonsupersymmetric vacua. A more detailed analysis will be
presented in \ISSn.

\subsec{Supersymmetry breaking by rank condition \ISS}

Our final example in this section is more complicated.  In
involves several fields transforming under a large symmetry group.
The fields $X_i$ in \worg\ are replaced by a matrix of fields.
Apart from the intrinsic interest in this example, it will also be
useful in our discussion in section 4.

Consider a theory with fields $\varphi$, $\tilde\varphi$, $\Phi$,
and parameters $f$, with global\foot{For our discussion in section 4,
we will take the $SU(n)$ symmetry to be gauged, but IR free.  In that
case, the $U(1)_R$ symmetry below is anomalous
(a linear combination of $U(1)_R$ and $U(1)_A$ is anomaly free, but broken by the parameter $f$), but is restored as an approximate, accidental
symmetry in the IR.  Also, the $SU(n)$ $D$-terms will vanish in the vacua. The results
discussed here will be completely unaffected by the weak gauging of $SU(n)$ in section 4.} symmetries
 \eqn\fieldsrank{\matrix{&&SU(n)&SU(N_f)_L&SU(N_f)_R
 &U(1)_V&U(1)_R&U(1)_A\cr
 \varphi &&{\bf n}& {\bf \overline N_f}& {\bf 1} &1&0 &1\cr
 \widetilde \varphi && \overline {\bf n}& {\bf 1 }&
 {\bf N_f} &-1& 0 &1\cr
 \Phi  &&{\bf 1 } &{\bf N_f}&{\bf \overline N_f}&0& 2 &-2\cr
 f &&{\bf 1 } &{\bf  \overline N_f}&{\bf N_f}&0& 0 &2\cr
 }}
We will take
 \eqn\nNf{n <N_f.}

 We take the K\"ahler potential $K$ to be canonical,
and the superpotential is
 \eqn\superrank{W= h \Tr\, \Phi \varphi \widetilde \varphi^T
 +  \Tr\, f \Phi,}
where $h$ is a coupling constant and the trace is over the global
symmetry indices.  The last term in \superrank\ respects the
symmetries in \fieldsrank\ because of the transformation laws of
the parameter $f$. Alternatively, the parameter $f$ breaks
$SU(N_f)\times SU(N_f)$ to a subgroup, and breaks $U(1)_A$, but it
does not break the $SU(n)$ symmetry or the R-symmetry.

Supersymmetry is broken when \nNf\ is satisfied. Consider the
$F$-component of $\Phi$
 \eqn\FPhi{-F^\dagger_\Phi = h  \varphi \widetilde \varphi^T +f}
(here we use $\dagger$ even in the classical theory because of the
flavor indices of $\Phi$).  This is an $N_f\times N_f$ matrix
relation. Because of \nNf, the first term is a matrix of rank $n$.
On the other hand, we can take $f$ to have rank larger than $n$,
up to rank $N_f$.   Therefore, if the rank of $f$ is larger than
$n$, and in particular if $f$ is proportional to the unit matrix
$\unit _{N_f}$,  then \FPhi\ cannot vanish,  $F_\Phi \neq 0$, and
supersymmetry is broken.

When \nNf\ is not satisfied, there are supersymmetric vacua, as in
the example \wlinq, which is similar to the case $n=N_f=1$.  The
difference is that, when \nNf\ is satisfied, there are not enough
additional degrees of freedom, $\varphi$ and $\tilde\varphi$, at
$\Phi =0$ to restore supersymmetry.

For simplicity, we take $f\equiv  -  h\mu^2 \unit _{N_f}$,
proportional to the unit matrix.  The minimum of the potential is
then at
 \eqn\Vrank{V= (N_f-n) |h\mu^2|^2}
and it occurs along the pseudomoduli space
 \eqn\Vmin{
 \Phi =\pmatrix{0&0 \cr 0& \Phi_0}, \qquad
 \varphi=\pmatrix{\varphi_0\cr
          0},
 \qquad \tilde \varphi =\pmatrix{\tilde \varphi_0\cr 0},\qquad
{\rm with}\,\,\,\,
\varphi_0 \tilde\varphi_0^T= \mu^2\unit_{n},
 }
and arbitrary $\Phi _0$, $\varphi_0$ and $\tilde \varphi_0$
(subject to the constraint in \Vmin). The first entries in \Vmin\
are the first $n$ components, and the second are the remaining
$N_f-n$ components, so e.g. $\Phi _0$ is a $(N_f-n)\times (N_f-n)$
square matrix.  The non-zero $F$ terms are $F_{\Phi _0}=\bar h
\bar \mu ^2\unit _{N_f-n}$.  The massless Goldstino comes from
the fermionic components of $\Phi _0$.

\subsec{One-loop lifting of pseudomoduli}

As we have seen in the examples above, models of tree-level
spontaneous supersymmetry breaking generally have classical moduli
spaces of degenerate, non-supersymmetric, vacua.  Indeed, the
massless Goldstino is in a chiral superfield (for $F$-term
breaking), whose scalar component is a classical pseudomodulus.
The example of section 2.3 shows that this is the case even if
this space of classical vacua becomes unstable in a region in
field space.  The example of section 2.7 \Vmin\ shows that there
can be additional pseudomoduli. We said above that we should use
the term ``pseudomoduli" space for the space of classical
non-supersymmetric vacua, because the degeneracy between these
vacua is usually lifted once quantum corrections are taken into
account.  In this section, we review how this comes about.

We will be interested in the one-loop effective potential (the
Coleman-Weinberg potential) for the pseudomoduli (such as $X$),
which comes from computing the one-loop correction to the vacuum
energy
 \eqn\CWgen{\eqalign{
 V^{(1)}_{eff} =& {1\over 64\pi^2}{\rm STr}\, \left( \CM^4\log
 {\CM^2\over M_{cutoff} ^2}\right)\cr
 \equiv &{1\over 64\pi ^2}\left[ \Tr \, \left( m_B^4 \log {m_B^2
 \over M_{cutoff} ^2} \right)
 -\Tr \, \left( m_F^4 \log {m_F^2\over M_{cutoff} ^2}\right)
 \right],}}
where $m_B^2$ and $m_F^2$ are the tree-level boson and fermion
masses, as a function of the expectation values of the
pseudomoduli, and $M_{cutoff}$ is a UV cutoff.  In \CWgen, ${\cal
M}^2$ stands for the classical mass-square matrix of the various
fields of the theory.

We would like to make two comments about the divergences in this
expression:
 \item{1.} In non-supersymmetric theories the effective potential
 includes also a quartic divergent term proportional to
 $M_{cutoff}^4\, {\rm STr}\, \unit $ and a quadratic divergent term
 proportional to $M_{cutoff}^2\, {\rm STr}\, \CM^2$.  They vanish in
 supersymmetric theories.
 \item{2.} The logarithmic divergent term $(\log M_{cutoff})\,  {\rm
 STr}\, \CM^4$ in \CWgen\ can be absorbed into the renormalization
 of the coupling constants appearing in the tree-level vacuum energy
 $V_0$ (see below). In particular, ${\rm STr}\,\CM^4$ is independent
 of the pseudomoduli.

For completeness, we recall the standard expressions for these
masses. For a general theory with $k$ chiral superfields,
$\Phi^a$, with canonical classical K\"ahler potential, $K=\Phi^a
\bar \Phi^a$, and superpotential $W(\Phi^a)$:
 \eqn\mbmf{m_0^2=\pmatrix{\bar W^{ac}W_{cb}&\bar W^{
 abc}W_c\cr W_{abc}\bar W^{c}&W_{ac}\bar W^{cb}},\qquad
 m_{1/2}^2=\pmatrix{\bar W^{ac}W_{cb}&0\cr 0&W_{ac}\bar W^{
 cb}},}
with $W_c\equiv \partial W/\partial Q^c$, etc., and $m_0^2$ and
$m_{1/2}^2$ are $2k\times 2k$ matrices.  Note that
 \eqn\strms{{\rm STr}\,\CM^2=0}

We will be interested in situations where we integrate out some
massive fields $\Phi^a$ whose superpotential is locally of the
form
 \eqn\effsup{W=\half \Phi^a M_{ab} \Phi^b +\dots,}
where $M_{ab}$ can depend on various massless fields $X$.
Integrating out $\Phi^a$ leads to the one loop effective K\"ahler
potential
 \eqn\keffgen{K_{eff}^{(1)}=-{1\over 32\pi ^2} \Tr [MM^\dagger
 \log (MM^\dagger/M_{cutoff} ^2)].}
If the supersymmetry breaking is small, we can use the effective
K\"ahler potential to find the effective potential.  For example,
if $M_{ab}$ depends on one pseudomodulus $X$, the effective
potential is
 \eqn\Vtruncs{V_{trunc}=(K_{eff\ X,\bar X})^{-1}|\partial_X W|^2.}
However, as we will discuss below, \Vtruncs\ gives the correct
expression for the effective potential \CWgen\ only to leading
order in $F_X= - {\overline{\partial_X W} \over K_{eff\ X, \bar
X}}  $. (It is verified in \ISS\ that \Vtruncs\ and \CWgen\ agree
to order ${\cal O}(F_X\bar F_X)$.)  Higher powers of $F_X$ arise
from terms in the low energy effective Lagrangian with more
superspace covariant derivatives, e.g.\ terms of the form
 \eqn\highdt{\int d^4\theta\, H(X,\bar X) ( D X)^2 + c.c.}
for some function $H(X, \bar X)$.  They cannot be ignored when the
supersymmetry breaking is large. The full effective potential
\CWgen\ includes all these higher order corrections.

\bigskip
\centerline{\it Example 1 -- the theory of section 2.3}

As a first application, we compute the one-loop potential on the
supersymmetry breaking pseudomoduli space mentioned in section
2.3. Recall that this space exists for $X$ outside of the range
\notatt\ where there is a tachyon, so we limit ourselves to
$|X|^2>|f/h|$. We treat the pseudomodulus $X$ as a background, and
use the masses \masspss\ in \CWgen.  This yields
 \eqn\Vonesim{\eqalign{
 V^{(1)}(|X|)= & {1 \over 64 \pi^2}\Big[ -2 |hf|^2\log
 M_{cutoff}^2 - 2|h X|^4 \log| h X|^2 \cr
 &+ ( |h X|^2- |hf| )^2
  \log( |h X|^2- |hf|) + ( |h X|^2+ |hf| )^2 \log(|h X|^2+ |hf|)
 \Big]\cr
 =&{|hf|^2\over 32 \pi^2}\left[ \log\left|{hX \over
 M_{cutoff}}\right|^2 +  {3\over 2} + v(z)  \right]\cr
 z\equiv &\left|{f \over h X^2}\right|\cr
 v(z)\equiv& {1\over 2} \left(z^{-2}(1+z)^2\log(1+z)+z^{-2}(1-z)^2
 \log(1-z)-3\right) =  -{z^2 \over 12 }+ \CO(z^4),}}
where the shift by $3\over 2$ is for later convenience.

The potential \Vonesim\ lifts the degeneracy along the
pseudomoduli space. It is an increasing function of $|X|$. It
pushes $X$ into the region \notatt; i.e.\ toward the region with a
tachyon (where the expression \Vonesim\ no longer makes sense).
From there, the theory falls into its supersymmetric vacua
\simpexsusy.

We will now use this simple example, and result \Vonesim, to
clarify and illustrate a number of technical points.  Similar
statements will apply to other examples.

Let us clarify the nature of the semiclassical limit. We take $h
\to 0$ (the coupling $h$ is IR free) with $f,\, X,\, q \sim
h^{-1}$ (and therefore $z \sim h^0$).  In this limit the classical
Lagrangian, based on canonical K\"ahler potential and the
superpotential \wlinq, scales like $h^{-2}$.   The one loop
corrections, in particular \Vonesim,  are of order $h^0$.  We can
neglect higher loop terms, which are order $h^2$ and higher.

Next, we want to understand the dependence on the UV cutoff
$M_{cutoff}$.  We define the running coupling
 \eqn\voorrrs{
 f(\mu) = f_{bare} \left(1+{|h^2|\over
 64 \pi^2}  \left({3\over 2} + \log{\mu^2\over M_{cutoff}^2 } \right)
 +\CO(h^4) \right),}
where we have set an additive constant to a convenient value.  In
terms of this running $f$ the potential \Vonesim\ is independent
of the UV cutoff $M_{cutoff}$
 \eqn\voxslbs{
 V(X)= |f(|hX|)|^2\left(1+{|h^2|\over 32\pi^2}  v(z)  + \CO(h^4)
 \right) .}
Here $f(\mu=|hX|)$ is the running coupling \voorrrs\ at the scale
of the massive fields $q$.

Equivalently, we can remember that in supersymmetric theories
there is only wavefunction renormalization.  The potential arises
from $F_X$, and therefore at the leading order only $Z_X$ can
affect the potential. The renormalization of $f$ in \voorrrs\ can
be understood as coming from $Z_X$, as
 \eqn\vzrens{V=Z_X^{-1}|\partial_X W|^2+{\rm finite}=
 Z_X^{-1}|f|^2 +{\rm finite}.}
We thus have
 \eqn\renconds{-{\partial V\over \partial \ln M_{cutoff} ^{2}}
 =\gamma _X|f|^2= {1\over 64\pi ^2}{\rm Str} \CM ^4
 + \CO(h^2),}
where we recognize $\gamma_X$ as the anomalous dimension of $X$.

A special situation arises when the supersymmetry breaking mass
splittings are effectively small.  This happens when $z\equiv
|f/hX^2| \ll 1$; i.e.\ either for small $|f|$, or for large $|X|$.
Expanding \Vonesim\ we find
 \eqn\Vonesime{
 V\approx|f|^2+ {|hf|^2\over 32 \pi^2}\left[ \log\left|{hX \over
 M_{cutoff}}\right|^2 +  {3\over 2} \right] + \CO(h^4) =
 |f(hX)|^2.}
This can be interpreted as arising from renormalization of the
K\"ahler potential
 \eqn\KXbarX{K_{ren}= |X |^2- {| h X |^2 \over 32 \pi^2}\left(
 \log\left|{ h X\over  M_{cutoff}}\right|^2  -{1\over 2} \right)
 + \CO(|h|^4).}
Note that this expression for the renormalized $K$ is valid also
for $f=0$, where supersymmetry is not broken along the moduli
space parameterized by $\ev{X}$.

We should also comment that since as $X\to 0$ the coupling
constant $h$ is renormalized to zero, the expression \KXbarX\
becomes accurate for small $X$ (though still outside of the
tachyonic range \notatt).

We have just seen that for small $z$ we can study a supersymmetric
low energy theory with superpotential $W=fX$ and an effective
K\"ahler potential given by \KXbarX.  This is a special case of
the discussion above about the K\"ahler potential \keffgen. Using
$M=h X$ in in \keffgen\ and $W=fX$, the approximate effective
potential \Vtruncs\ agrees with \Vonesime.

As discussed around \Vtruncs, the supersymmetric effective
potential \Vtruncs\ is valid only when the supersymmetry breaking
is small. The correct one-loop effective potential is given by
\CWgen\ (which in our simple example is given by \Vonesim),
whether or not the supersymmetry breaking is small.  In general,
additional contributions which are not included in \Vtruncs\ are
higher orders in $|f|$ in \Vonesim\ (i.e. the function $v(z)$ in
\Vonesim).

\bigskip
\centerline{\it Example 2 -- the basic O'Raifeartaigh model
(section 2.5)}

We now compute the one loop correction to the pseudomodulus
potential in the O'Raifeartaigh model, example 1 of section 2.5.
The classical flat direction of the classical pseudomodulus $X$ is
lifted by a quantum effective potential, $V_{eff}(X)$ \HuqUE.

We again treat the pseudomodulus $X$ as a background. The one-loop
effective potential $V_{eff}(X)$ is given by the expression
\CWgen, using the classical masses \eigenei\ and \eigenii.  As
follows from the R-symmetry, $V_{eff}(X)$ depends only on $|X|$.
We find that the potential $V_{eff}(X)$ is a monotonically
increasing function of $|X|$, with the following asymptotic
behavior at small and large $|X|$:
 \eqn\voxsmalllarge{
 V_{eff}(X)=\cases{
     V_0 + m_X^2 |X|^2 + \CO(|X|^4)  & $X\approx 0$ \cr
     |f|^2\left(1 + \gamma_X \left(\log \left|{hX \over M_{cutoff}}
     \right|^2 + {3\over 2}\right) +\CO(h^4, {\log |X|\over |X|^4})
    \right)& $ X\to \infty $}}
where the constants are
 \eqn\voxcon{\eqalign{
 V_0=& |f|^2\left[1+{|h^2|\over 32\pi^2} \left(
 \log{|m|^2\over M_{cutoff}^2 }+  {3\over 2} + v(y)  \right) +
 \CO(h^4)  \right]\cr
 y=& \left| {hf\over m^2}\right|\cr
 v(y)=& {1\over 2} \left(y^{-2}(1+y)^2\log(1+y)+y^{-2}(1-y)^2
 \log(1-y)-3\right) =  -{y^2 \over 12 }+ \CO(y^4) \cr
 m_X^2=& {1\over 32\pi ^2}\left|{h^4f^2\over m^2}\right| \nu(y)
 + \CO(h^4) \cr
 \nu(y)=& y^{-3}\left((1+y)^2\log (1+y)-(1-y)^2\log
 (1-y)-2y\right) = {2  \over 3} + \CO(y^2) \cr
 \gamma_X=& {|h|^2 \over 32 \pi^2} + \CO(h^4)
 .}}
The function $v(y)$ is as in \Vonesim\ but its argument here, $y$,
depends only on the coupling constants, and is independent of the
pseudomodulus $X$. Recall that we take the parameter $y$, defined
in \ydef,  to be in the range $0\leq y\leq 1$.

As in the previous example, the semiclassical limit is $h \to 0$
(the coupling $h$ is IR free) with $f,\, X,\, \phi_{1,2} \sim
h^{-1}$ and $m \sim h^0$ (and therefore $y \sim h^0$).

Also, as in that example, the running coupling constant
 \eqn\voorrr{
 f(\mu) = f_{bare} \left(1+{|h^2|\over
 64 \pi^2}  \left({3\over 2} + \log{\mu^2\over M_{cutoff}^2 } \right)
 +\CO(h^4) \right),}
removes the dependence on the UV cutoff $M_{cutoff}$
 \eqn\voxslb{\eqalign{
 V(x)=&\cases{
     V_0 + m_X^2 |X|^2 + \CO(|X|^4)  & $X\approx 0$ \cr
     |f(hX)|^2+... & $ X\to \infty $}
     \cr
 V_0=&|f(m)|^2\left(1+{|h^2|\over 32\pi^2}  v(y)  + \CO(h^4)
 \right) .}}

Let us discuss the effective potential in the two limits $X
\approx 0$ and $|X| \to \infty$.  The sign of the mass square in
\voxcon\ is positive, signaling that the potential has a minimum
at $X=0$. The behavior for large $X$ is dominated by the
renormalization group running of the effective coupling constant
at the scale $|h X|$, which is the scale of the masses in the
problem.  Finally, it is easy to show using the full expression
from \CWgen\ that the one loop potential is monotonic between
these two limits, and therefore $X=0$ is the global minimum of the
potential.

Again, as in the previous example, for $y\equiv |hf/m^2| \ll 1$,
the supersymmetry breaking is small.  Then, the effective
potential can alternatively be computed in the supersymmetric
low-energy effective theory, with $K$ given by \keffgen\ and
$W=fX$, leading to the effective potential \Vtruncs.  The potential
\CWgen\ applies more generally.

For example, expanding around the minimum at $X=0$, \Vtruncs\ only
reproduces the leading order term in the expansion in $y \ll 1$
for $m_X^2$ in \voxcon. It fails to reproduce the answer for
larger values of $y$, e.g.
 \eqn\mxyi{m_X^2={|h^3 f|\over 16\pi ^2}(\log 4-1)
 \qquad\hbox{for}\qquad |hf|=|m|^2 \qquad ; \qquad y=1.}
On the other hand, even if $y$ is not small, the higher order $F$ terms are
insignificant far from the origin of  the pseudomoduli space, and indeed there the
truncated potential \Vtruncs\ agrees with the full effective
potential \voxsmalllarge:
 \eqn\vonefar{V^{(1)}\to \gamma _X^{(1)}\log\left({|hX|^2\over
 M_{cutoff} ^2}\right)|f|^2 \qquad \hbox{for $hX$ large}.}

Let us now consider the modified model of section 2.6, where we
add $\half h \epsilon \phi_2^2$ to the superpotential \worm. As we
saw, there are then two supersymmetric states at \SUSYmo, and
there can also be a metastable state near $X=0$. Including the
$\epsilon$ correction to the mass eigenvalues, the one-loop
potential \CWgen\ now has a linear term in $X$ (a tadpole) at
$X=0$, with coefficient $\CO (\epsilon)$.  The quadratic term in
$X$ is not much changed by the $O(\epsilon)$ correction, so the
upshot is a local minimum of the one-loop potential at $X\sim
\epsilon$.

To summarize this example, we found in section 2.6 that the theory
with nonzero $f$ and $\epsilon$ has a classical pseudomoduli space
of nonsupersymmetric vacua, which is sensible in the range
\nontapm\ (which includes the region around $X=0$), where there
are no tachyonic modes. Now we have shown that the one-loop
effective potential lifts this pseudomoduli space, and stabilizes
$X$ near the origin.  For $\epsilon \ll 1$, the tachyonic
direction down to the supersymmetric vacua \SUSYmo\ only appears
at large $X$, so the metastable vacuum near the origin, with
broken supersymmetry, can be parametrically long lived.

It is straightforward to repeat the  computation of the one-loop
effective potential for the model where supersymmetry is broken by
the rank condition (section 2.7).  Again, we set $f= -h\mu^2
\unit$, and then we find that most of the degeneracy along the
classical pseudomoduli space \Vmin\ is removed by the one-loop
effective potential \CWgen.  The masses of the fluctuations of
$\Phi$, $\varphi$ and $\tilde \varphi$, as a function of the
pseudomoduli in \Vmin, are found to be similar to those of the
O'Raifeartaigh model given in \eigenei\ and \eigenii, with
$m^2=hf\equiv -h \mu ^2$ (so $y=1$ in \ydef).  The $SU(n)$ gauge
fields do not contribute to \CWgen, since their spectrum is
supersymmetric to this order. Up to symmetry transformations, the
vacua are found to be at
 \eqn\Vminq{
 \Phi =\pmatrix{0&0 \cr 0& 0}, \qquad
 \varphi= \tilde \varphi =\pmatrix{\mu \unit_n \cr 0}.
 }
The vacua \Vminq\ spontaneously break the global symmetry, $G\to
H$. Associated with that, the vacua \Vminq\ actually form a
compact moduli space of vacua, ${\cal M}_{vac}=G/H$, parameterized
by the massless Goldstone bosons.  Since this space of vacua is
associated with an exact global symmetry breaking it is robust,
and the degeneracy is not lifted by higher order corrections.  In
particular, these vacua cannot become tachyonic. The one-loop
potential computed from \CWgen\ gives non-tachyonic masses to {\it
all} other pseudomoduli,  so the vacua \Vminq\ are true local
minima of the effective potential \ISS.

\subsec{Relation to R-symmetry \NelsonNF}

Consider a generic theory and ask for a condition for broken
supersymmetry.  This means that we cannot solve all the equations
 \eqn\feqns{\partial_a W(\Phi ) =0
 \qquad\hbox{for all $a=1\dots k$.}}
But if $W$ is a generic superpotential, then \feqns\ involves $k$
equations for the $k$ quantities $\Phi ^a$, so generally they can
all be solved.  Non-R flavor symmetries do not help.  Consider for
example a global non-R $U(1)$ symmetry.  Then, the equations
\feqns\ can be written as $k-1$ independent equations for $k-1$
independent unknowns, as seen by writing
 \eqn\kmoc{W=W(t^a=\Phi ^{a}\Phi _1^{-q_a/q_1}) \qquad\hbox{
 $a=2\dots k$}. }
($q_a$ is the $U(1)$ charge of $\Phi^a$).  But if there is an
R-symmetry, then we can write
 \eqn\kcond{W=T f(t^a= \Phi ^a \Phi
 _1^{-r_a/r_1})\qquad\hbox{ $T= \Phi_1^{2/r_1}$}, }
($r_a$ is the R-charge of $\Phi^a$), and then in terms of $T$ and
$t^a$ for generic $f$ the equations \feqns\ set $T=0$ which is a
singular point.  Away from $T=0$ the equations are
over-constrained: they are $k$ equations for $k-1$ independent
unknowns, so generically they cannot be solved.  Exceptions occur
either for a non-generic $f$, or when a solution with $T=0$ and
therefore $\Phi_1=0$ is allowed.  This is the case when $r_1=2$
and all other $r_a=0$. Then there is a $k-2$ dimensional space of
supersymmetric vacua, at $\Phi _1=0$, $f(\Phi _a)=0$. (More
generally, there are exceptional cases with supersymmetry unbroken
for fields at the origin, when all fields, for which the K\"ahler
potential is smooth, have non-negative R-charges less than 2.)

These observations about the relation between R-symmetry and
supersymmetry breaking fit with the examples above.

The simplest theory (section 2.1) with $W=fX$ has an R-symmetry
and broken supersymmetry.  Adding e.g.\ $\Delta W=\half \epsilon
X^2$ breaks the R-symmetry, and restores supersymmetry.

This is also true for its generalization with more complicated $K$
of section 2.2, which depends only on $X\bar X$.  If $K$ depends
separately on $X$ and $\bar X$ (not only through the combination
$X\bar X$), the theory does not have an R-symmetry but
supersymmetry is still broken. This shows that we can have broken
supersymmetry without R-symmetry.  Here it happens because the
superpotential is not a generic function of $X$.

The addition of light fields as in section 2.3 preserves the
R-symmetry, but restores supersymmetry. This demonstrates that
having an R-symmetry does not guarantee that supersymmetry is
broken.  This example realizes the exceptional case, $r_1=2$,
$r_{a\not=1}=0$, mentioned above.

The example of section 2.4 has a $U(1)_R$ symmetry, and indeed
there is no static supersymmetric vacuum.  But there is a runaway
direction, along which supersymmetry is asymptotically restored.
This illustrates the need to still check for runaway directions.

The O'Raifeartaigh type models of section 2.5 have an R-symmetry,
and broken supersymmetry for generic $g_1(\phi)$ and $g_2(\phi)$.
The example 2 there, with non-generic $g_1(\phi)$ and $g_2(\phi)$,
illustrates that having an R-symmetry does not guarantee broken
symmetry, if the superpotential is not generic.

The deformation \worm\ of the O'Raifeartaigh model in section 2.6
breaks the R-symmetry, and indeed restores supersymmetry. However,
for small $\epsilon$ there is an approximate R-symmetry which is
related to supersymmetry breaking in the metastable state.

Finally, the models based on the rank condition of section 2.7
have an R-symmetry and correspondingly they have broken
supersymmetry, for $n<N_f$.  (For $n\geq N_f$, supersymmetry
is not broken, by a generalization of the comment following \kcond\
about the case $r_1=2$, with all other $r_a=0$.) As mentioned in footnote 6,
we will later discuss this model with the $SU(n)$ symmetry gauged, but IR free.
The $U(1)_R$ symmetry is then only an approximate symmetry.
Correspondingly, the supersymmetry breaking (with $n<N_f$) will be
in metastable vacua \ISS.

To summarize, generically there is broken supersymmetry if and
only if there is an R-symmetry.  There is broken supersymmetry in
a metastable state if and only if there is an approximate
R-symmetry.  For realistic models of supersymmetry breaking, we
need to break the R-symmetry, to get gaugino masses.  To avoid
having a massless R-axion if the symmetry is spontaneously broken
it should also be explicitly broken.  Gravity effects can help
\BaggerHH, but ignoring gravity, we conclude that realistic and
generic models of supersymmetry breaking require that we live in a
metastable state.

\newsec{Supersymmetric QCD}

In this section we will discuss the dynamics of supersymmetric QCD
(SQCD) for various numbers of colors and flavors.  This section
will be brief.  We refer the reader to the books and reviews of
the subject, e.g.\ \refs{\TerningBQ,\Dinebook, \ISrev\PeskinQI
\ShifmanUA - \StrasslerQG}, for more details.

\subsec{Super Yang-Mills theory -- $N_f=0$}

A pure gauge theory is characterized by a scale $\Lambda$.  At
energy of order $\Lambda$, it confines and leads to nonzero gluino
condensation, breaking a discrete R-symmetry.

For $SU(N_c)$ gauge theory we define the gauge invariant chiral
operator
 \eqn\Sdef{S\equiv -{1\over 32\pi ^2}\Tr\, W^\alpha W_\alpha=
 {1\over 32\pi ^2} \Tr\, \left( \lambda \lambda + ...+
 \theta\theta ( \half F^{\mu\nu} F_{\mu\nu}+ ...) \right), }
which can be interpreted as a ``glueball'' superfield.  Here we
follow the Wess and Bagger notation \WessCP\ where $\lambda
\lambda \equiv \lambda^\alpha \lambda_\alpha$.  The dynamics leads
to gaugino condensation:
 \eqn\glucond{\ev{S}={1\over 32\pi ^2}\ev{\Tr\, \lambda \lambda }
 =(\Lambda^{3N_c})^{1 \over N_c}}
where branches of the fractional power in \glucond\ represent the
values in the $N_c$ different supersymmetric vacua. The theory has
an anomaly free  ${\bf Z}_{2N_c}$ discrete symmetry (left unbroken
by instantons), and \glucond\ implies that it is spontaneously
broken to ${\bf Z}_2$.

The $N_c$ supersymmetric vacua with \glucond\ are those counted by
the Witten index, $\Tr (-1)^F=N_c$ \WittenDF.  Since $\lambda
\lambda$ is the first component of the chiral superfield $S$, the
expectation values \glucond\ do not break supersymmetry.

The relation \glucond\ is exact.  This can be seen by promoting
$\Lambda$ to an expectation value of a background chiral
superfield \refs{\SeibergVC,\IntriligatorJR}, which is assigned
charge $R(\Lambda)=2/3$ to account for the anomaly.  There is no
correction to \glucond\ compatible with this $R$ charge assignment
and holomorphy\foot{The non-zero value of the coefficient in
\glucond\ can be set to one in a particular renormalization
scheme. See \FinnellDR\ for discussion, and comparison with
various instanton calculations.}.

The gaugino condensation  can be represented as a nontrivial
superpotential
 \eqn\symew{ W_{eff} = N_c( \Lambda^{3N_c})^{1 \over N_c}. }
Comments:
 \item{1.} The superpotential \symew\ is independent of fields.
 It is meaningful when coupling to supergravity, or if $\Lambda$
 is a background field source.
 \item{2.} Equation \symew\ can be used to find the tension
 of domain walls interpolating between these vacua labelled by
 $k_1$ and $k_2$ \DvaliXE
 \eqn\tensionkokt{T_{k_1,k_2}=\left| N_c( \Lambda^{3N_c})^{1
  \over N_c} (e^{2\pi i k_1 \over N_c} - e^{2\pi i k_2 \over
  N_c})\right|.}
 \item{3.} Thinking of $3N_c\log \Lambda $ as a source for the
 operator $S \sim \Tr W_\alpha^2$ we can find
  \eqn\vevS{\ev{S}={1 \over 3N_c} \partial_{\log \Lambda} W_{eff} =
  (\Lambda^{3N_c})^{1 \over N_c}.}
 \item{4.} Using this observation we can perform a Legendre
 transform to derive the Veneziano-Yankielowicz superpotential
 \VenezianoAH
  \eqn\WofS{ W_{eff}(S)= N_c S(1 - \log S/\Lambda^3) .}
 It should be stressed that $S$ is not a light fields and
 therefore this expression is not a term in the Wilsonian
 effective action.  It is a term in the 1PI action and therefore
 it can be used only to find $\ev{S}$ and tensions of domain
 walls.  However, there is no particle-like
 excitation (e.g.\ a glueball) which is described by the field $S$.

\subsec{Semiclassical SQCD}

We consider $SU(N_c)$ gauge theory with $N_f$ quarks $Q$ and $N_f$
anti-quarks $\tilde Q$.

The gauge and global symmetries are
 \eqn\symms{\matrix{
 &&SU(N_c)&[SU(N_f)_L&SU(N_f)_R&U(1)_B&U(1)_R&U(1)_A]\cr
 Q&&{\bf N_c}&{\bf N_f}& {\bf  1} &1&1-{N_c\over N_f}&1\cr
 \widetilde Q&&{\bf \overline N_c} &{\bf  1} &{\bf \bar N_f}
 &-1&1-{N_c\over N_f}&1\cr
 }}
Here the global symmetries are denoted by $[...]$.  The $U(1)_A$
symmetry is anomalous and the other symmetries are anomaly free.
We also assign charges to the coupling constants: regarding them
as background chiral superfields leads to useful selection rules
\SeibergVC, \eqn\csymms{\matrix{
 &&SU(N_c)&[SU(N_f)_L&SU(N_f)_R&U(1)_B&U(1)_R&U(1)_A]\cr
 m&&{\bf 1}&{\bf \bar N_f}& {\bf  N_f} &0&2{N_c\over N_f}&-2\cr
 \Lambda ^{3N_c-N_c}&&{\bf 1} &{\bf  1} &{\bf 1}
 &0&0&2N_f\cr
 }}
Here $m$ is a possible mass term that we can add, $W_{tree}=\Tr\,
m\tilde Q Q$, and $\Lambda$ is the dynamical scale, related to the
running gauge coupling as
 \eqn\rengrc{\Lambda^{3N_c-N_f} = e^{-{8\pi^2 \over g^2(\mu)} + i
 \theta } \mu^{3N_c-N_f}.}
Instanton  amplitudes come with the factor of
$\Lambda^{3N_c-N_f}$, and their violation of the $U(1)_A$ symmetry
is accounted for by the charge assignment in \csymms.

As seen from \rengrc, the theory is UV free for $N_f <3 N_c$, i.e.
$g^2(\mu )\to 0$ for $\mu \gg |\Lambda|$. On the other hand, for
$N_f\geq 3N_c$, the theory is IR free, i.e. $g^2(\mu )\to 0$ for
$\mu \ll |\Lambda|$ (for $N_f=3N_c$ the beta function vanishes at
one loop, but at two loops it is IR free).

In the rest of this subsection, we take $W_{tree}=0$. The
classical potential is then
 \eqn\Dterme{V \sim \sum_a (D^a)^2 = \sum_a (\Tr (Q T^a
 Q^\dagger - \tilde Q^*T^a \tilde Q^T))^2}
($T^a$ are the $SU(N_c)$ generators).  It leads to flat directions
which we refer to as the classical moduli space of vacua
$\CM_{cl}$.  As is always the case, $\CM _{cl}$ can be understood
in terms of gauge invariant monomials of the chiral superfields,
and the light moduli in $\CM _{cl}$ can be understood as the
chiral superfields that are left uneaten by the Higgs mechanism.

For $N_f<N_c$ up to gauge and flavor rotations, ${\cal M}_{cl}$ is
given by \AffleckMK
 \eqn\Qvevs{Q=\widetilde Q= \pmatrix { a_1& & & & \cr
 &a_2& & & \cr & & . & & \cr & & & a_{N_f} &\quad \cr}.}
Its complex dimension is ${\rm dim}_{\bf C}{\cal M}_{cl}=N_f^2$.
The gauge invariant description is ${\cal M}_{cl}=\{ M^f_{\tilde
g}=(\tilde Q Q^T)^f_{\tilde g}\}$, $f, \tilde g=1\dots N_f$. The
gauge group is broken on $\CM _{cl}$ as $SU(N_c)\rightarrow
SU(N_c-N_f)$.  The classical K\"ahler potential on $\CM _{cl}$ is
 \eqn\clasKnfncl{K_{cl}= 2\, \Tr\, \sqrt{M^\dagger M}.}
(To see that, write the D-term equations as $ Q^\dagger Q=\tilde
Q^T \tilde Q^*$, and use it find $M^\dagger M = Q^* \tilde
Q^\dagger \tilde Q Q^T= (Q^* Q^T)^2$.  Then the K\"ahler potential
is $\tr\, Q^\dagger Q + \tr\, \tilde Q^\dagger \tilde Q=2\, \tr\,
\sqrt{M^\dagger M}$.) This is singular near the origin. As always,
singularities in the low-energy effective theory signal new light
fields, which should be included for a smooth description of the
physics.  Here the singularities of $K_{cl}$ occur at subspaces
where some of the $SU(N_c)/SU(N_c-N_f)$ gauge bosons become
massless, and they need to be included in the description.

For $N_f\geq N_c$ we have  ${\rm dim}_{\bf C}{\cal
M}_{cl}=2N_cN_f-(N_c^2-1)$. Up to gauge and flavor rotations
\AffleckMK,
 \eqn\Qvevl{Q=
 \pmatrix { a_1& & & \cr &a_2& & \cr & & .  & \cr & & & a_{N_c}\cr
 & & & \cr & & & \cr} , \quad \widetilde Q= \pmatrix { \widetilde
 a_1& &       &         \cr
   &\widetilde a_2&     &         \cr
   &   & .   &         \cr
   &   &     & \widetilde a_{N_c}\cr
   &   &     &         \cr
   &   &     &         \cr}, \qquad
 |a_i|^2 - |\widetilde a_i|^2= {\rm independent ~ of} ~ i.}
The gauge invariant description is given by the fields $M=\tilde Q
Q^T$, $B=Q^{N_c}$ (contracted with the epsilon-symbol), $\tilde
B={\tilde Q}^{N_c}$,
subject to various classical relations,
\eqn\nfgncrlns{{\cal M}_{cl}=\{M, \ B,
 \tilde B | \ C_i(M, B, \tilde B)=0\}.}
The functions $C_i$, giving the classical relations,  are of
course compatible with the symmetries \symms, including $U(1)_A$.
For example for $N_f=N_c$, we have \SeibergBZ
 \eqn\nfnccoa{{\cal M}_{cl}=\{M_{\tilde g}^f, \ B,
 \tilde B | \det M -B\tilde B=0\},}
where the constraint follows from $\det M =\det Q \det \tilde
Q=B\tilde B$.  The spaces \nfgncrlns, for all $N_f\geq N_c$,  are
singular at the origin, $M=B=\tilde B=0$, because it is possible
to set all $C_i=0$, and also all variations $\delta C_i=0$ there.
The classical interpretation is that the $SU(N_c)$ gauge fields,
which are massless at the origin, need to be included for the
low-energy effective theory to be non-singular.

For $N_f>N_c$, among other constraints, the $N_f\times N_f$ matrix
$M=\tilde Q Q^T$
 satisfies
  \eqn\rankm{{\rm rank}(M)\leq N_c \qquad\hbox{classically}.}

\subsec{Adding large quark mass terms}

Consider adding quark masses, via the tree-level superpotential
 \eqn\massterm{W_{tree}=\Tr\, m \tilde Q Q^T\equiv  \Tr\, mM .}
For large $m$ (more precisely, the eigenvalues of $m$ are much
larger than $|\Lambda|$) we can integrate out the quarks and the
low energy theory is a pure gauge theory. Its scale $\Lambda_L$ is
determined at one loop as
 \eqn\matchingr{\Lambda_L^{3N_c}= \det m \ \Lambda^{3N_c-N_f}.}
Gluino condensation in this theory leads, as in \symew, to
 \eqn\glc{W_{eff} = N_c (\det m\  \Lambda^{3N_c-N_f})^{1 \over
 N_c};}
it follows from holomorphy and symmetries that \glc\ is the exact
effective superpotential. The superpotential \glc\ can be
interpreted as part of the generating functional for correlation
functions, with the mass $m$ in \massterm\ acting as the source
for the operator $M$, and $\log \Lambda ^{3N_c-N_f}$ as the source
for the operator $S\sim \Tr W_\alpha W^\alpha$
\refs{\IntriligatorJR, \IntriligatorUK}. We can thus use \glc\ to
find
 \eqn\Mvev{\eqalign{
 &\ev{M}_{susy} = \partial_m W_{eff}= (\det m\
 \Lambda^{3N_c-N_f})^{1 \over N_c} {1 \over m} \cr
 &\ev{S}_{susy}= \partial_{\log \Lambda ^{3N_c-N_f}} W_{eff}= (\det
 m\ \Lambda^{3N_c-N_f})^{1 \over N_c}.
 }}
The subscript emphasizes that these are the expectation values in
the supersymmetric vacua.  Note that there are $N_c$ solutions in
\Mvev, differing by a $N_c$-th root of unity phase, which
correspond to the $\Tr (-1)^F=N_c$ supersymmetric vacua of the
low-energy super-Yang-Mills theory. The result \Mvev\ is valid for
all $N_f$.  It is interesting to note that, for $N_f>N_c$, the
matrix $\ev{M}$ in \Mvev\ does not satisfy the classical
constraint \rankm\ of the theory with massless flavors; however,
taking $m\to 0$ in \Mvev\ does bring $\ev{M}$ back to ${\cal
M}_{cl}$.

Performing a Legendre transform between $m$ and $M$, we can use
\glc\ to derive the 1PI effective action
 \eqn\WofM{ W_{eff}(M)= (N_c-N_f) \left({\Lambda ^{3N_c-N_f}\over
 \det M}\right)^{1/(N_c-N_f)} + \Tr mM.}
One might be tempted to interpret \WofM\ also as a Wilsonian
effective action for the light field $M$.  However, as we will
discuss below, this is not always correct.

Finally we can introduce the field $S$ into \WofM\ by performing a
Legendre transform with respect to its source $\log
\Lambda^{3N_c-N_f}$ to find \TaylorBP\
 \eqn\WofMS{ W_{eff}(M,S)= S \left((N_c-N_f)- \log {S^{N_c-N_f}
 \det M \over \Lambda^{3N_c-N_f}} \right)  + \Tr mM.}
Again, this expression can be used to find the expectation values
\Mvev\ and to study domain wall tensions, but it should not be
viewed as a term in a Wilsonian effective action.

\subsec{$N_f< N_c$ massless flavors \AffleckMK}

We have seen that the classical theory has a moduli space of
supersymmetric vacua $\CM_{cl}$.  We now explore the low energy
effective Lagrangian along $\CM_{cl}$ and examine whether a
superpotential can be generated there.  The symmetries \symms\
constrain the superpotential to be of the form \DavisMZ
 \eqn\wdynis{W_{dyn}\propto
 \left({\Lambda ^{3N_c-N_f}\over \det M}\right)^{1/(N_c-N_f)}.}
Therefore, we face a dynamical question of determining the
coefficient in \wdynis.  Note that \wdynis\ is non-perturbative,
because of the positive power of $\Lambda \sim \exp(-8\pi^2/(3N_c
-N_f) g^2)$.

Recall that the gauge group is Higgsed to $SU(N_c-N_f)$ on the
classical moduli space. For $N_f=N_c-1$, the gauge group is
completely Higgsed, and then there are finite action (constrained)
instantons which generate \wdynis.  For $N_f<N_c-1$, \wdynis\ is
instead associated with gaugino condensation in the unbroken
$SU(N_c-N_f)$ -- that is the reason for the fractional power in
\wdynis. Finally, comparing with \WofM\ we see that the
coefficient in \wdynis\ must be $N_c-N_f$
  \eqn\wdyniss{W_{dyn}= (N_c-N_f)
 \left({\Lambda ^{3N_c-N_f}\over \det M}\right)^{1/(N_c-N_f)}.}

For $N_f\geq N_c$,  \wdynis\ does not make sense.  For $N_f=N_c$,
the exponent diverges.  For $N_f>N_c$, the constraint \rankm\
implies $\det M =0$. Therefore,  for $N_f\geq N_c$ massless
flavors, the quantum theory has a moduli space of inequivalent
vacua.

\subsec{$N_f= N_c$ massless flavors \SeibergBZ}

Here the vacuum degeneracy cannot be lifted by $W_{dyn}$, so the
moduli space is still parameterized by the gauge invariant fields
$M$, $B$ and $\tilde B$.  But the classical constraint \nfnccoa\
they satisfy is modified (consistent with the symmetries \symms\
and \csymms)
 \eqn\nfnccoq{{\cal M}_{qu}=\{M_{\tilde g}^f, \ B,
 \tilde B | \det M -B\tilde B=\Lambda^{2N_c}\}.}
Note that this is a nonperturbative effect, proportional to a
positive power of $\Lambda$.  So, as is appropriate, the
deformation is important only near the origin, and is negligible
at large fields, relative to $\Lambda$,  where the theory is
weakly coupled. Indeed, the power in \nfnccoq\ is precisely that
associated with a one instanton correction to the constraint in
\nfnccoa. The constraint \nfnccoq\ can be seen from \Mvev, which
for $N_f=N_c$ has $\det M=\Lambda ^{2N_c}$, independent of $m$.
(One can introduce sources for the operators $B$ and $\tilde B$,
to get the full constraint \nfnccoq.)  The space ${\cal M}_{cl}$
in \nfnccoa\ was singular at $M=B=\tilde B=0$, but the space
\nfnccoq\ is everywhere smooth.  The only light degrees of freedom
of the low-energy effective theory are the moduli of \nfnccoq.

The theory with the modified constraint can be described using a
Lagrange multiplier $X$ and a superpotential
 \eqn\nfncmc{W=X(\det M - B\tilde B - \Lambda^{2N_c}),}
but it should be stressed that this is not a term in a Wilsonian
action.  There is no light field $X$ and similarly, the mode of
$M$, $B$ and $\tilde B$ which is proportional to $\det M - B\tilde
B $ are not light.  However, \nfncmc\ is still a useful way to
implement the constraint.

\subsec{$N_f > N_c$ \NSd}

The vacuum degeneracy of the theory with massless flavors again
cannot be lifted by $W_{dyn}$. Moreover, for all $N_f>N_c$, the
classical moduli space constraints \nfgncrlns\ cannot be deformed
because no deformation would be compatible with holomorphy and the
symmetries in \symms\ and \csymms.  So there is a quantum moduli
space of vacua, coinciding with the classical moduli space
\nfgncrlns, ${\cal M}_q={\cal M}_{cl}$.  The singularity of these
spaces at the origin indicates additional, massless degrees of
freedom there.  Their nature is clarified by a duality.

The original $SU(N_c)$ theory, with $N_f$ flavors, is dual to
another gauge theory based on the gauge group $SU(n=N_f-N_c)$ with
spectrum of fields and couplings
\eqn\magns{\matrix{&&SU(n)&[SU(N_f)_L&SU(N_f)_R
 &U(1)_B&U(1)_R&U(1)_A]\cr
 \varphi &&{\bf n}& {\bf \bar N_f}& {\bf 1} &{N_c\over n} &1-{n\over
 N_f} &1\cr
 \widetilde \varphi && \overline {\bf n}& {\bf 1 }&
 {\bf  N_f} &-{N_c\over n}&1-{n\over N_f} &1\cr
 \Phi  &&{\bf 1 } &{\bf  N_f}&{\bf\bar N_f}&0& 2{n\over N_f}
 &-2\cr
 f &&{\bf 1 } &{\bf \bar  N_f}&{\bf N_f}&0& 2-2{n\over N_f} &2\cr
\Lambda ^{3n-N_f}&&{\bf 1}&{\bf 1}&{\bf 1}&0&0&2N_f\cr
 }}
(again, the group in $[...]$ is a global symmetry) with canonical
$K$ for the fields $\varphi$, $\widetilde\varphi$, and $\Phi$, and
superpotential
 \eqn\superranks{W= h\, \Tr\ \Phi \varphi \widetilde \varphi^T+\Tr\
 f\Phi.}
As we will discuss, the coupling $f$ is proportional to the mass
of the electric quarks. In particular, if $m=0$ in the electric
theory, then $f=0$ in the magnetic theory.  $U(1)_A$ in \magns\ is
anomalous but the other symmetries are not. The scale $\tilde
\Lambda$ of the magnetic theory can be taken to be the same as the
$\Lambda$ of the electric theory, as we indicate in \magns.

We refer to the original theory \symms\ as electric and to \magns\
as magnetic. This duality between the electric and the magnetic
theories states that these two different theories have the same IR
behavior. Better agreement between the two theories is obtained if
we modify the K\"ahler potential by higher order terms.

Comments:
 \item{1.} The anomaly free symmetries of the electric and the
 magnetic theories are the same.  All 'tHooft anomaly matching
 conditions of these symmetries are satisfied.
  \item{2.} The relations between the variables of the electric
  and magnetic descriptions are
  \eqn\emrel{M= \tilde Q Q^T = \alpha \Lambda \Phi \qquad, \qquad
  B=Q^{N_c}  = \beta^n \Lambda ^{2N_c-N_f}\varphi^n}
 with some dimensionless constants $\alpha$ and $\beta$. (Below we
 will determine $\alpha$.) It is easy to check that the
 identification of operators \emrel\ is consistent with
 the anomaly free symmetries.  (An alternative description was
 given in \ISrev, where the scales of the electric and magnetic
 theories were taken to be different; the descriptions are
 equivalent, as reviewed, e.g.\ in \ISS.)
 \item{3.} For ${3\over 2}N_c <N_f < 3N_c$, the electric
 and magnetic theories are both UV free, and they differ in the UV.
The two different UV free starting points flow under the
renormalization group (RG) to the same interacting RG fixed point
in the IR. A detailed discussion of this RG flow can be found,
e.g.\ in \StrasslerQG.
 \item{4.} For $N_c+2 \le N_f \leq {3\over 2}N_c$ the magnetic
theory is IR free, with irrelevant interactions. The UV free
electric theory flows at long distance to the IR free magnetic
theory.
 \item{5.} For $N_f=N_c+1$ we can still use the variables in
\magns\ but without the magnetic gauge fields and with the
addition of a term proportional to $\det \Phi$ to the
superpotential \SeibergBZ.
  \item{6.} Turning on mass terms $\Tr\, m Q\tilde Q = \Tr\, m M$
  in the electric theory is described by adding to the magnetic
  superpotential $\Lambda \alpha \Tr\, m \Phi$.  We will analyze
  it in detail in the next subsection.

\subsec{Adding small mass terms}

We again add \massterm
  \eqn\massterma{W_{tree}=\Tr\, m \tilde Q Q^T =\Tr\, m M}
but this time we take the masses (eigenvalues of $m$) small
compared with $|\Lambda|$. Now, we should be able to reproduce the
expectation values \Mvev\ from our low energy effective theory.

For $N_f<N_c$, the low energy theory has
$W_{exact}=W_{dyn}+W_{tree}$, which gives precisely the
superpotential \WofM.  The Legendre transform in \Mvev\ ensures
that setting $F_M^\dagger=- \partial _M W_{exact}=0$ yields the
$N_c$ supersymmetric vacua at $\ev{M}$ given in \Mvev.

As we mentioned above, for $N_f\geq N_c$, \WofM\ is not meaningful
as a superpotential on the moduli space.  Rather, it should be
viewed as a superpotential on a larger field space, where $M$ is
arbitrary rather than subject to \rankm, and which is meaningful
only for nonzero $m$.  As we are going to discuss, the dual theory
provides an interpretation of this.

For $N_f=N_c$ \WofM\ does not make sense.  Instead, we can find
$\ev{M}$ using the superpotential \nfncmc.

For $N_f=N_c+1$ we have to add \WofM\ to the superpotential (as
commented after \emrel).

For $N_f>N_c+1$ the meaning of \WofM\ is slightly more subtle.
Consider moving the field $\Phi \sim M$ away from its expectation
value.  The superpotential \superranks\ gives masses to the dual
quarks $\varphi$.  Using an expression like \symew\ for gluino
condensation in the magnetic gauge group leads to
 \eqn\mglucond{W=n (h^{N_f} \det \Phi \Lambda^{3n-N_f})^{1\over n}.}
where we set the scales of the magnetic and electric theories to
be the same $\Lambda$.  This agrees with \WofM\ provided
 \eqn\gccond{ h^{N_f} \det \Phi \Lambda^{3n-N_f} =
 (-1)^{N_f-N_c} {\det M \over \Lambda^{3N_c-N_f}}}
which fixes the coefficient $\alpha$ in \emrel\
 \eqn\fixal{M=(-1)^{1 - {N_c\over N_f}} h \Lambda \Phi.}
Correspondingly, the coefficient $f$ in \superranks\ is related to
the electric mass by
 \eqn\mfreln{f=\alpha \Lambda m=(-1)^{1+{N_c\over N_f}}mh\Lambda.}

\newsec{Dynamical supersymmetry breaking}

We will now consider four typical examples of DSB.  The common
feature of these examples is that at low energies they can be
given a semiclassical supersymmetric description as in the
examples in section 2. The first three examples which are based on
the dynamics of $N_f < N_c$, $N_f=N_c$ and $N_f >N_c$ were found
in the 80s, 90s and 00s respectively. The fourth example, which is
based on the dynamics of $N_f=0$, allows us to easily convert any
example in section 2 to a model of DSB.

Many other examples of DSB are known. Some of them are strongly
coupled and do not admit a semiclassical supersymmetric
description involving an effective K\"ahler potential and an
effective superpotential (examples are $SU(5)$ or $SO(10)$ gauge
theories with a single generation of quarks and leptons
\refs{\AffleckVC,\MeuriceAI}). In other situations the
question of supersymmetry breaking is inconclusive (e.g.\ an
$SU(2)$ gauge theory with matter in the four dimensional
representation \IntriligatorRX). In addition, many variants of the
examples below are known and they exhibit various interesting
features (see, e.g.\
\refs{\AffleckUZ\AffleckXZ\MurayamaNG\PoppitzWP \CsakiGR
\IntriligatorFK\MurayamaPB \DimopoulosWW\LutyNY\DimopoulosJE-\LutyNQ }).
Additional review and references can be found in e.g.\
\refs{\ShadmiJY, \TerningTH, \TerningBQ, \Dinebook}

\subsec{The (3,2) model \AffleckXZ}

The gauge group is
 \eqn\thrtg{SU(3) \times SU(2)}
and we have chiral superfields: $Q$ in $({\bf 3,2})$, $\tilde u$
in $({\bf \bar 3,1})$, $\tilde d$ in $({\bf \bar 3,1})$, $L$ in
$({\bf 1,2})$. For $W_{tree}=0$, the classical moduli space is
given by arbitrary expectation values of the gauge invariants
 \eqn\thrtgi{X _1= Q \tilde d L\qquad ,  \qquad X_2 = Q \tilde
u L \qquad, \qquad Z=QQ\tilde u \tilde d.}  Both gauge groups are
Higgsed on this classical moduli space. We add to the model a tree
level superpotential
 \eqn\ttwt{W_{tree} = \lambda Q \tilde d L = \lambda X _1.}
This theory has a $U(1)_R$ symmetry, with $R(Q)=-1$, $R(\tilde
u)=R(\tilde d)=0$, $R(L)=3$. A crucial aspect of \ttwt\ is that it
lifts all of the classical D-flat directions. Therefore, the
theory does not have any runaway directions.

Using the global symmetries (including those under which the couplings,
treated as background chiral superfields, are charged), the exact
superpotential for the fields \thrtgi\ is
 \eqn\thrtdy{W_{exact} = {\Lambda_3^7 \over Z}+\lambda X_1.}
The first term in \thrtdy\ is $W_{dyn}$, which is generated by an
$SU(3)$ instanton. This theory dynamically breaks
supersymmetry\foot{A quick way to see that is to note that
$W_{dyn}$ pushes $Z$ away from the origin, which spontaneously
breaks the $U(1)_R$ symmetry.  There is thus a compact moduli
space of vacua, whose modulus is the massless Goldstone boson.  If
supersymmetry were unbroken, the Goldstone boson would have a
scalar superpartner, which would lead to a non-compact moduli
space - but that cannot be the case, because $W_{tree}$ lifts all
of the classical flat directions \AffleckVC.}.

For $\lambda \ll 1$, the vacuum is at large expectation value for
the fields. Since the gauge groups are Higgsed at a high energy
scale, their running coupling is weak. Because the theory is
weakly coupled for the fields in this limit, we have  $K\approx
K_{classical}$, so the K\"ahler potential is under control. It is
then easy to find that the field expectation values and the vacuum
energy density at the minimum are
 \eqn\approxa{ v\sim \Lambda _3/\lambda ^{1/7} \qquad ;
 \qquad V=M_S^4 \sim | \lambda ^{10/7}\Lambda _3^4|}
(the precise coefficient can be computed, using $K=K_{cl}$). Note
that, to justify $K\approx K_{cl}$, we need $v\gg \Lambda _3$ and
also $v\gg \Lambda _2$, and the latter condition requires $\Lambda
_3\gg \lambda ^{1/7}\Lambda _2$. In addition to the massless
Goldstino, there is a massless Goldstone boson, because the vacuum
spontaneously breaks the $U(1)_R$ symmetry.

The above analysis is valid when $\Lambda _3\gg \Lambda _2$. As
seen from the expressions above, in this limit the $SU(2)$ gauge
dynamics scale $\Lambda _2$ does not appear directly in the
approximate answers \approxa. The $SU(2)$ gauge group is weakly
coupled at the scale $\Lambda _3$, and the role of the $SU(2)$
gauge symmetry is simply to restrict the possible superpotential
couplings, and its classical gauge potential lifts certain
directions in field space thus avoiding runaway. The fact that
$\Lambda _2$ does not enter into \thrtdy\ fits with the fact that
the $SU(2)$ gauge group has $N_f=N_c$. So, as reviewed in section
3.5, it does not contribute to $W_{dyn}$, but instead leads to the
quantum modified moduli space constraint \SeibergBZ\ of \nfnccoq.
The quantum modified moduli space is neglected in the analysis
above, and that is justified when $\Lambda _3\gg \Lambda _2$.

On the other hand, in the limit $\Lambda _2\gg \Lambda _3$, the
$SU(2)$ group becomes strong first in the RG flow to the IR, and
it is then essential to include the quantum modified moduli space
constraint.  Below the scale $\Lambda _2$, the light fields are
$q=QL/\Lambda_2$, in the ${\bf 3}$ of $SU(3)$, and $\tilde
q=Q^2/\Lambda _2$, and $\tilde u$ and $\tilde d$, all in the ${\bf
\bar 3}$, subject to the quantum constraint $q\tilde q=\Lambda
_2^2$. The constraint breaks $SU(3)$ to $SU(2)'\subset SU(3)$, at
the scale $\Lambda _2$, and $q$ and $\tilde q$ are Higgsed. The
fields $\tilde u$ and $\tilde d$ each decompose as ${\bf 3, \bar
3\to 2+1}$ under $SU(3)\to SU(2)'$, so we have $SU(2)'$ with
$N_f=1$ flavor, plus two singlets.  In the limit, we obtain a
superpotential which is similar to \thrtdy, but with a different
interpretation of the terms. In particular, the $\lambda X_1$ term
is interpreted as $\lambda \Lambda _2^2 S_d$, where $S_d$ is the
$SU(2)'$  singlet from $\tilde d$.  In the $\lambda ^{1/7}\Lambda
_2\gg \Lambda _3$ limit, the $SU(2)'\subset SU(3)$ dynamics is
insignificant, and we have $M_S^4=\alpha |\lambda ^2\Lambda
_2^4|$, where $\alpha$ is a positive ${\cal O}(1)$ K\"aher
potential coefficient, $K\supset {1\over \alpha}S_d\bar S_d$ that
cannot be directly calculated \IT.

\subsec{Modified moduli space example \refs{\IT,\IY}}

Consider the $SU(N_c)$ theory with $N_f=N_c$ and add fields
$S_a^{\tilde a}$, $b$ and $\tilde b$ and a superpotential (up to
coupling constants)
 \eqn\ITtr{W_{tree} =\,\tr\, S \tilde Q
 Q^T + b \det \tilde Q + \tilde b \det Q.}
Classically $Q=\tilde Q=0$.  In the quantum theory we get the
effective superpotential (see \nfncmc)
 \eqn\ITef{W_{effective} =\, \tr\, S M + b
 \tilde B + \tilde b  B + X(\det M - B \tilde B -\Lambda^{2N_c})}
which breaks SUSY.  This breaking is dynamical.  It depends on the
IR confinement of the $N_f=N_c$ theory, from quarks and gluons in
the UV, into the composite fields $M$ and $B$ and $\tilde B$ in
the IR and on the quantum deformation of the moduli space by
$\Lambda^{2N_c}$ in \nfnccoq.

Let us specialize to $N_f=N_c=2$, where the fundamentals and
anti-fundamentals can be written as $2N_f=4$ fundamentals
$Q^{fc}$, $f=1\dots 4$, $c=1,2$.  The gauge invariants are
$U^{fg}= Q^{fc}Q^{gd}\epsilon _{cd}$, in the ${\bf 6}$ of the
global $SU(4)\cong SO(6)$ flavor symmetry.  To emphasize that it
is an $SO(6)$ vector we will also express it as
 \eqn\Visvec{\vec V=(V^1={1 \over 2} (U^{12} + U^{34}),
 V^2={i \over 2}(U^{12} - U^{34}),...).}
The quantum moduli space constraint \nfncmc\ for this case is
\SeibergBZ\
 \eqn\qms{{\rm Pf}\, U= U^{12} U^{34} - U^{13} U^{24}+
 U^{14}U^{23} = \vec V\cdot \vec V = \Lambda ^4.}
We add singlets $\vec S$, also in the ${\bf 6}$ of the global
flavor $SO(6)$, with superpotential
 \eqn\wqms{W_{tree}={1 \over 2} h S_{fg} Q^{fc}Q^{gd}\epsilon
 _{cd}= 2 h \vec S\cdot \vec V,}
where $S_{fg}$ is related to $\vec S$ as in \Visvec\ and the
factor of $2$ arises from this change of notation. Unlike
\ITtr\ITef, here we have explicitly exhibited the coupling
constant $ h $. There is a conserved $U(1)_R$ symmetry, with
$R(Q)=0$, and hence $R(\vec V)=0$, and $R(\vec S)=2$. Because
$\bar F_{\vec S} =- 2h \vec V$, the constraint \qms\ implies that
$F_{\vec S}\neq 0$, so SUSY is broken.

Let us analyze it in more detail.  We start with the classical
theory.  The superpotential coupling $\half h  S_{fg}
Q^{fc}Q^{gd}\epsilon _{cd}$ lifts all the flat directions with
nonzero $Q$.  So the classical moduli space is the space of $\vec
S$.  Moving far out along these flat directions the fundamental
quarks are massive and can be integrated out.  The low energy
$SU(2)$ gauge theory has scale $\Lambda_L^6=\Lambda^4 h^2 \vec
S\cdot \vec S$, and its gluino condensation generates
 \eqn\wlowintoq{W_{low}=2(\Lambda_L^6)^{1/2}= 2
  \left(h^2 \Lambda^4 \vec S \cdot \vec S\right)^{1\over 2}.}
Using the symmetries and holomorphy it is easy to see that
\wlowintoq\ is exact.  Now it is clear that for any nonzero $\vec
S$ the superpotential is not stationary, and the point $\vec S=0$
is singular and needs to be examined in detail.

Before we conclude that supersymmetry is broken away from the
origin we have to examine the potential at infinity to make sure
that there is no runaway. Using the classical K\"ahler potential
for $\vec S$ which is canonical, the superpotential \wlowintoq\
leads to
 \eqn\VclIT{V_{cl}=4 |h\Lambda^2|^2 {\vec S \cdot \bar
 {\vec S} \over |\vec S \cdot \vec S|}.}
Depending on the direction in the space this expression either
diverges at infinity or asymptotes to a constant
$4|h\Lambda^2|^2$.  It is straightforward to include the one loop
correction to this expression.  This situation is very similar to
the discussion around \Vonesim.  The fundamental quarks $Q$ are
massive and their loop leads to logarithmic corrections to the
potential which makes it grow at infinity.  We conclude that the
pseudoflat directions with broken supersymmetry in \VclIT\ is
lifted and pushes the system to smaller values of $\vec S$.

When $|h\vec S| \ll |\Lambda|$ the superpotential \wqms\ gives the
quarks small masses and they cannot be integrated out so easily.
But then we can use our understanding of the macroscopic theory,
where the $SU(2)$ gauge fields and matter of the microscopic
theory are replaced in the IR with the fields $\vec V$, subject to
the constraint \qms.   We solve this constraint as
 \eqn\vecVe{\vec V =\Lambda (\sqrt{\Lambda ^2-\vec v^2}, \vec v),}
where $\vec v$ is an $SO(5)$ vector.  We will assume that $|\vec
v|\ll |\Lambda|$.  This assumption is valid up to symmetry
transformations near the origin of the classical theory, where we
expect to find our ground state. Similarly, we write $\vec S\equiv
(S_1, \vec s)$, where $\vec s$ is an $SO(5)$ vector. Then \wqms\
is
 \eqn\wqmsii{W= 2 h  \Lambda S_1\sqrt{\Lambda ^2-\vec
 v^2}+ 2 h  \Lambda \vec v\cdot \vec s \approx 2 h  \Lambda^2 S_1 -
  h S_1\vec v^2+2 h  \Lambda \vec v\cdot \vec s  .}
 The K\"ahler potential for the fields $S_1$, $\vec s$, and $\vec v$
 is smooth, and can be taken to be
 \eqn\kmacroex{K=S_1\bar S_1 +\vec s \cdot\bar{\vec s}
 +{1\over \alpha} \vec v\cdot \bar {\vec v}
 +{\cal O}({1\over |\Lambda |^2}),}
where $\alpha$ is an ${\cal O}(1)$ coefficient that we cannot
determine.

Up to symmetry transformations, the vacua have arbitrary
$\ev{S_1}$, and $\vec v=\vec s=0$.  This leads to a seven real
dimensional pseudomoduli space.  Its dimensions include the two
non-compact directions given by $\ev{S_1}$, and five real
Goldstone bosons living on $SO(6)/SO(5)\cong S^5$, coming from
components of $\vec v$ and $\vec s$.

We can integrate out the massive modes of $\vec v$ to find an
effective superpotential.  For $\vec s=0$ it is $W_{eff}=2 h \Lambda^2
S_1$, and more generally, it is given by $W_{eff}=2 \left(h^2 \Lambda^4
\vec S \cdot \vec S\right)^{1\over 2}$ which agrees with
\wlowintoq.

Supersymmetry is broken by $-\bar F_{S_1}=2 h  \Lambda ^2\neq 0$.
Since $F_{S_1}$ is generated by dimensional transmutation, the
supersymmetry breaking is dynamical.  The massless Goldstino comes
from $S_1$.

We should now examine how this pseudomoduli space is lifted in the
quantum theory.  This is easily done using the low energy theory
based on the superpotential \wqmsii\ and the K\"ahler potential
\kmacroex\ by noticing that it is a multi-field analog of the
$y=1$ O'Raifeartaigh model.  The one-loop potential \CWgen\ lifts
the degeneracy and leads to a supersymmetry breaking minimum at
$\vec S=0$ \ChackoSI.  At this vacuum the global $SO(6)$ symmetry
is spontaneously broken to $SO(5)$ by the constraint \qms, but the $U(1)_R$ symmetry is
unbroken.  So there is a five real dimensional, compact space of
supersymmetry breaking vacua, given by the Goldstone boson
manifold $SO(6)/SO(5)\cong S^5$.

For $ h  \ll 1$, we can have large $S_1$ and still use the low
energy effective theory provided
 \eqn\macroval{ | h  S_1| \ll |\Lambda| \ll |S_1|.}
In this limit,  the behavior of the one-loop potential \CWgen,
computed in the low-energy effective field theory, asymptotes as
in \vonefar\ to
 \eqn\vonefarIT{V^{(1)}\to \gamma ^{(1)}_{macro}
 \log \left({|2 h  S_1|^2\over M_{cutoff}^2}\right)
 |2 h  \Lambda ^2|^2.}
As we have reviewed, the dependence on $M_{cutoff}$ can be
absorbed into the renormalization of $ h $.  The coefficient in
\vonefarIT\ is the anomalous dimension of the pseudomodulus,
computed in the macroscopic theory.  It depends on the ${\cal
O}(1)$ unknown constant $\alpha$ in \kmacroex.  Since $\gamma
_{macro}^{(1)}>0$, the potential \vonefarIT\ is an increasing
function of $|S_1|$.

On the other hand, as we remarked above, if $|\Lambda| \ll | h
S_1|$, then, we should instead use the microscopic theory.  The
result for the potential is similar to \vonefarIT, though with a
different, but again positive, numerical coefficient $\gamma
^{(1)}_{micro}$ for the one-loop anomalous dimension of $S_1$,
computed from the microscopic $Q$ fields running in the loop
\ArkaniHamedUT.  We cannot compute the potential in the
intermediate range, $| h S_1|\sim |\Lambda|$, but in all
calculable regions the potential slopes toward the origin,
$S_1=0$.

\bigskip
\centerline{\it Deforming the model}

Consider adding a $U(1)_R$ breaking, but $SO(6)$ invariant, term
 \eqn\ITdef{\Delta W= \half \epsilon \vec S^2}
to \wqms.  Adding this to \wlowintoq\ or
\wqmsii, the theory has a five complex dimensional, non-compact,
moduli space of supersymmetric vacua
 \eqn\susITm{\vec S= -{2h \over \epsilon} \vec V \qquad ;
 \qquad \vec V^2= \Lambda ^4.}
For $|\epsilon |\gg |\Lambda|$, the fields $\vec S$ are heavy and
can be integrated out. The low energy theory is simply the $SU(2)$
theory with four massless doublets and no superpotential (the
cubic couplings of \wqms\ do not lead to a quartic superpotential
when $\vec S$ is integrated out).  This has a moduli space which
is reproduced by \susITm.

For $|\epsilon| \ll |\Lambda|$,  the $\vec S$ fields are light, and
need to be included in the low energy theory;  i.e.\ we add
\ITdef\ to \wqmsii.  As we take $\epsilon \to 0$, the SUSY vacua \susITm\
run off to infinity. In addition to these supersymmetric ground
states at large $|\vec S|$, we still have the compact moduli space of supersymmetry
breaking vacua discussed following \wqmsii, with $\vec S$ near the origin. For $|\epsilon |\ll  |\Lambda |$
these metastable, supersymmetry breaking states are very long lived. Finally, as $\epsilon
\to 0$ the supersymmetric states disappear from the Hilbert space
and we are left with only the metastable states.

Note that these theories provide examples of nonchiral theories
that dynamically break supersymmetry. How is that compatible with
the Witten index \WittenDF?  The argument based on the Witten
index relies on adding mass terms to the theory and tracking the
supersymmetric states as the mass is removed.  In this problem we
can add two possible mass terms. First, we can add mass terms for
the fundamental quarks.  This is done in the effective theory by
adding $\vec m \cdot \vec V$ to the superpotential. But this has
no effect because $\vec m$ can be absorbed in a shift of $\vec S$.
Second, if we add \ITdef, $\vec S$ is massive.  For large mass it
leads to the non-compact moduli space of supersymmetric states
\susITm.  For small mass we also find the compact moduli space of
supersymmetry breaking metastable states, and as $\epsilon \to 0$
the supersymmetric states disappear from the Hilbert space and
supersymmetry is broken.

\subsec{Metastable states in SQCD \ISS}

Consider SQCD with $N_c+1 \le N_f <{3 \over 2}N_c$, with small quark
masses
 \eqn\smallm{|{\rm Eigenvalues}( m)| \ll |\Lambda|.}
The range of $N_f$ is such that the magnetic dual \NSd\ of section
3.6 is the IR free, low-energy effective field theory. We thus
analyze the groundstates in the magnetic dual, with superpotential
 \eqn\isssu{h \Tr\, \Phi \varphi \tilde\varphi + \alpha \Lambda
 \Tr\, m\Phi.}
This is the same as the theory we studied in
\fieldsrank\superrank\ with the identification\foot{The global
vector $U(1)$ symmetry in \fieldsrank\ is normalized differently
than the baryon number symmetry in \magns. Also, the $U(1)_R$
symmetry in \magns\ is anomaly free but it is broken by the mass
term, while in \fieldsrank\ we took $U(1)_R$ to preserve the term
linear in $\Phi$ but it is anomalous.}
 \eqn\ideniss{\alpha\Lambda m =f .}
For simplicity, we will take $m$ (and therefore also $f$) to be
proportional to the unit matrix, thus preserving the global
$SU(N_f)$.

As discussed following \fieldsrank, this low energy theory has a
supersymmetry breaking minimum \Vminq.  All non-Goldstone modes
have non-tachyonic masses there, from the one-loop potential,
which is computed via \CWgen\ in the low-energy dual theory. The
fact that the magnetic theory is IR free ensures that higher loops
are suppressed, and in particular cannot invalidate the results
from the one-loop potential.

We thus conclude that SQCD has metastable dynamical supersymmetry
breaking vacua. In terms of the microscopic electric SQCD theory,
the DSB vacua \Vminq\ have zero expectation value for the meson
fields, $\ev{M}=0$, and non-zero expectation value of some baryon
fields, $\ev{B}\neq 0$ and $\ev{\tilde B}\neq 0$, which follow
from the non-zero $\ev{\varphi}$ and $\ev{\tilde\varphi}$ in
\Vminq. In terms of the IR dual magnetic theory, these vacua are
semi-classical, but in terms of the microscopic, electric SQCD
they are not, they are strongly quantum-mechanical.

As noted after \Vminq, the supersymmetry breaking vacua \Vminq\
spontaneously break the global symmetries, from $G=SU(N_f)\times
U(1)_B$ to $H=SU(N_f-N_c)\times SU(N_c) \times U(1)$.  Associated
with that, there is a compact moduli space of vacua, the manifold
of massless Goldstone bosons\foot{In various generalizations of
this example, these compact moduli spaces of DSB vacua can support
topological solitons, which can be (meta) stable, see \EtoYV\ for
a fuller discussion.}, ${\cal M}_{vac}=G/H$. Note that the DSB
vacua have an assortment of massless fields: the $G/H$ Goldstone
bosons and a number of massless fermions including the Goldstino,
which come from the fermionic components of the fields $\Phi _0$
in \Vmin. This is to be contrasted with the naive expectation that
there should be no massless fields (and, in particular, no
candidate Goldstino for DSB to occur), since the quarks $Q$ all
have a mass $m$, and the low-energy SYM gets a mass gap.  The dual
magnetic theory shows that this naive expectation is incorrect.

SQCD also has $N_c$ supersymmetric vacua, with mass gap and
$\ev{M}\sim \ev{\Phi}\neq 0$, and $\ev{B}=\ev{\tilde B}=0$. These
supersymmetric vacua arise from the effective interaction
\mglucond\ which, as explained earlier, are obtained from gluino
condensation in the magnetic theory.  Thus, in terms of the
magnetic dual theory, supersymmetry is non-perturbatively
restored, in a theory that breaks supersymmetry at tree-level.
Indeed, from the point of view of the theory
\fieldsrank\superrank, the R-symmetry is anomalous and is
explicitly broken (this is manifest with the interaction
\mglucond), and therefore supersymmetry is restored. As long as
$N_f$ is in the free magnetic range, $N_f<{3\over 2}N_c$, the
supersymmetry restoring interaction \mglucond\ is irrelevant at
the DSB vacua near $\Phi =0$.  Then the DSB and the SUSY vacua are
sufficiently separated for the DSB vacua to be meaningful.

The small mass condition \smallm\ has the following useful
consequences:
 \item{1.} It ensures that the analysis within
the low-energy effective field theory (the magnetic dual) is
valid: the superpotential coupling $f\sim m\Lambda $ is then
safely below the UV cutoff, $ \Lambda$, of the magnetic dual
theory.
 \item{2.} It ensures that effects from the microscopic (electric)
theory do not invalidate the macroscopic analysis of supersymmetry
breaking and the one loop stabilization of the vacua \Vminq.  A
way to see this is to note that the one-loop potential gives all
(non-Goldstone) pseudomoduli mass squares of order $|f| \sim
|m\Lambda|$ (much as in \mxyi) which is non-analytic in the
superpotential coupling $f \sim m\Lambda$.  This reflects the fact
that it comes from integrating out modes which become massless in
this limit. On the other hand, any effects from the microscopic
theory must be analytic in $m$, and then \smallm\ ensures that
such effects are subleading to \mxyi.
 \item{3.} The condition \smallm\ also ensures that the
supersymmetric vacua \Mvev\ can be seen in the magnetic effective
theory, as then \Mvev\ is safely below its cutoff, $|\ev{M}|\ll
|\Lambda|$.

\item{4.}  It ensures that the metastable state is parametrically
long lived. The tunneling probability is $\sim \exp(-S_{bounce})$,
where $S_{bounce}\sim \Delta \Phi^4/V_{meta}$, with $\Delta \Phi$
the separation in field space between the metastable and the
supersymmetric vacua, and $V_{meta}=M_s^4$. For small masses
\smallm, $S_{bounce}$ is parametrically large, and thus the
metastable DSB vacua can be made parametrically arbitrarily long
lived.

This kind of DSB appears generic.  It exists also in similar
$SO(N_c)$ and $SP(N_c)$ gauge theories \ISS, and many
generalizations of it were found recently (see e.g.\
\refs{\FrancoES\OoguriPJ\KitanoWM\KitanoWZ\AmaritiVK
\DineGM\DineXT\KitanoXG\MurayamaYF \AharonyMY- \CsakiWI}). Also,
the early universe favors populating the DSB vacua over the SUSY
vacua. One reason for that is the large degeneracy of the
Goldstone boson moduli space of DSB vacua, versus the discrete
$N_c$ mass gapped supersymmetric vacua.  Another reason is that
the DSB vacua are closer to the origin of the moduli space than
the supersymmetric vacua, and that is favored by the thermal
effective potential \refs{\AbelCR\CraigKX\FischlerXH-\AbelMY}.

\subsec{Naturalizing (retrofitting) models \refs{\DineGM,\DineXT}}

As we stressed in the introduction (around equation \scaleofs), in
order for a model of supersymmetry breaking to be fully natural,
all scales which are much smaller than the UV cutoff $M_{cutoff}$
should arise via dimensional transmutation.   To be fully natural,
the Lagrangian cannot have any super-renormalizable (relevant)
operators, since they are naturally of order a positive power of
$M_{cutoff}$.  The Lagrangian should have only renormalizable
(marginal) operators and non-renormalizable (irrelevant)
operators, which are suppressed by inverse powers of $M_{cutoff}$.
Any needed relevant operators should then arise dynamically, with
exponentially suppressed coefficients, as in \scaleofs.

A simple way to achieve that is the following. Consider an
``unnatural model'' of supersymmetry breaking like one of the
models in section 2, with superpotential terms like
$W_{tree}\supset f\CO_1+ m \CO_2$, where $\CO_1$ is some dimension
one operator, $\CO _2$ is a dimension two operator, and $f\equiv
\mu ^2$.  We want the mass scales $m$ and $\mu$ to be much less
than $M_{cutoff}$. Such a model can easily be naturalized (or
retrofitted) by removing these couplings from the theory and
replacing them with interactions with the operator $S\equiv -
\Tr\, W_\alpha^2 /32\pi ^2$ of some added, but otherwise
decoupled, pure Yang-Mills theory (with no charged matter):
 \eqn\naturater{\int d^2 \theta \left[-{8\pi ^2\over
 g^2(M_{cutoff})} + {a_1 \over M_{cutoff}}\CO_1 +
 {a_2 \over M_{cutoff}^2}\CO_2\right] S ,}
where $a_{1,2}$ are dimensionless coefficients of order one,
so the couplings in \naturater\ are natural.

The pure Yang-Mills theory entering in \naturater\ has a
dynamically generated scale $\Lambda $, which satisfies  $\Lambda
\ll M_{cutoff}$, as in \scaleofs. For energies below the scale
$\Lambda $, the added Yang-Mills theory becomes strong and leads
to gaugino condensation $\ev{S} = \Lambda ^3$. Substituting this
in \naturater\ we find
 \eqn\naturaterg{\int d^2 \theta \left[{a_1 \Lambda ^3\over
 M_{cutoff}}\CO_1  +
 {a_2 \Lambda ^3\over M_{cutoff}^2}\CO_2 \right].}
Thus we generate super-renormalizable couplings in the
superpotential with $\mu^2 \sim \Lambda^3/ M_{cutoff} \ll
M_{cutoff}^2$ and $m \sim \Lambda ^3/ M_{cutoff}^2\ll M_{cutoff}
$. For example, the O'Raifeartaigh model of section 2.5 can be
naturalized by replacing \wor\ with
 \eqn\naturex{\int d^2\theta \left[\half h X \phi _1^2+
 \left(-{8\pi ^2\over  g^2(M_{cutoff})} + {a_1 \over  M_{cutoff}}X
 + {a_2 \over M_{cutoff}^2}\phi _1\phi _2\right)S \right].}

More generally, we can use couplings like \naturater\ with
different gauge groups or with couplings with higher powers of
$W_\alpha$. This way, every unnatural model can be easily
naturalized.

This naturalization procedure is not unique.  A given macroscopic
theory can be naturalized in more than one way.  Consider, for
example, the macroscopic models based on the rank condition of
section 2.6. One way to naturalize them is to replace the last
term in \superrank\ with ${1\over M_{cutoff}} \Tr\, \Phi\, \Tr\,
W_\alpha'{}^2 $, where $W_\alpha '$ is the field strength of some
other pure Yang-Mills theory, with scale $\Lambda '$; this leads
to $f \sim \Lambda'{}^3/M_{cutoff}$. Alternatively, we can first
view this theory as the low energy approximation of a SQCD theory,
as in section 4.3.  This theory is not yet fully natural because
of the existence of the quark mass term $m \Tr\, \tilde Q Q^T$ in
the Lagrangian.  As in \ideniss, this leads to $f\sim m\Lambda$,
which is dynamical, but not yet fully natural because we need
\smallm, $|m|\ll |\Lambda |\ll M_{cutoff}$.  It can be made fully
natural by replacing the mass term of the UV lagrangian with ${1
\over M_{cutoff}^2} \Tr\, \tilde Q Q^T\, \Tr\, W_\alpha '{}^2$
\AharonyMY.  This leads to $m\sim \Lambda '{}^3/M_{cutoff}^2$, so
$|m|\ll |\Lambda|$ is natural, and $f\sim \Lambda \Lambda
'{}^3/M_{cutoff}^2$.

Throughout this analysis, we have viewed the theory in an
expansion in powers of $M_{cutoff}^{-1}$. For example, in
\naturex\ we did not consider higher dimension operators like
${X^2 \over M_{cutoff}^2} W_\alpha^2$. As another example, gluino
condensation in \naturex\ does not simply replace $\left(-{8\pi^2
\over g^2} +{X \over M_{cutoff} } \right)S$ with ${X \over
M_{cutoff} } \Lambda^3$.  More precisely, following the analysis
in section 3.1, for an $SU(N_c)$ gauge theory it replaces it with
 \eqn\naturep{N_c \Lambda^3\exp\left( {X\over N_c M_{cutoff} }
 \right) \approx N_c \Lambda^3 + {X \over M_{cutoff} } \Lambda^3,}
where we neglected higher order terms in $M_{cutoff}^{-1}$ in the
latter expression.

This expansion in powers of $M_{cutoff}^{-1}$ is significant.  It
is well known that one can trigger supersymmetry breaking by
coupling a chiral superfield to a Yang-Mills theory via higher
dimension operators and using gluino condensation
\refs{\FerraraQS\DerendingerKK-\DineRZ}.  This usually leads to
runaway behavior, as is clear from the first expression in
\naturep. However, since we content ourselves with finding
supersymmetry breaking only in a metastable state, we can focus on
a particular region in field space and ignore possible vacua
elsewhere in field space. This focusing on a region in field space
is achieved by the expansion in $M_{cutoff}^{-1}$ we mentioned
above. Therefore, this naturalization procedure leads to
acceptable, metastable, dynamical supersymmetry breaking.

\bigskip\bigskip\bigskip\bigskip\bigskip

\centerline{\it Acknowledgments}
\bigskip
We would like to thank the organizers of the various schools, and
also the participants for their questions and comments. We thank
our many colleagues and friends for useful discussions about these
topics.  In particular, we would like to thank our collaborators
on these and related subjects: I.~Affleck, M.~Dine, R.~Leigh,
A.~Nelson, P.~Pouliot, S.~Shenker, D.~Shih, M.~Strassler,
S.~Thomas and E.~Witten. The research of NS is supported in part
by DOE grant DE-FG02-90ER40542.  The research of KI is supported
in part by UCSD grant DOE-FG03-97ER40546.

\listrefs
\end